\documentstyle[aps,multicol,amssymb,graphicx]{revtex}

\begin{document}
\title{What should a statistical mechanics satisfy to reflect nature? \thanks{To appear in a special issue of Physica D entitled {\it Anomalous
Distributions, Nonlinear Dynamics, and Nonextensivity}, eds. H.L. Swinney and C. Tsallis.}}
\author{Constantino Tsallis\thanks{tsallis@cbpf.br}
}
\address{Centro Brasileiro de Pesquisas F\'{\i}sicas\\
Rua Xavier Sigaud 150, 
22290-180 Rio de Janeiro-RJ, Brazil.}
\maketitle

\begin{abstract}
There is no compelling reason imposing that the methods of statistical mechanics should be restricted to the dynamical systems which follow the usual Boltzmann-Gibbs prescriptions. More specifically, ubiquitous natural and artificial systems exhibit complex dynamics, for instance, generic stationary states which are {\it not} ergodic nor close to it, in any geometrically simple subset of the 
{\it a priori} allowed phase space, in any (even extended) trivial sense. A vast class of such systems appears, nevertheless, to be tractable within thermostatistical methods completely analogous to the usual ones. The question posed in the title arises then naturally. Some answer to this complex question is advanced in the present review of nonextensive statistical mechanics and its recent connections.
\end{abstract}


\begin{multicols}{2}

A basic answer to the question of the title could of course be ``to make predictions that are experimentally confirmed''. Still, what we would really like to know is what {\it a priori} mathematical properties or features should a formalism satisfy in order to enable useful statistical predictions in a manner similar to that of Boltzmann-Gibbs (BG) statistical mechanics. This task is either impossible or of extremely high difficulty. In order to be able to formulate an answer which could at least be partially satisfactory, let us analyze the basic structure of the BG theory.
Indeed, no logical-deductive path ever existed for proposing a new physical theory, or for generalizing a pre-existing one. In fact, such proposal frequently --- perhaps always --- occurs on a metaphorical basis. Therefore, the analysis of the structure of the BG theory will hopefully provide us a {\it metaphor} for formulating a statistical mechanics which could perhaps be more powerful than the one we already have thanks to the genius of Boltzmann, Gibbs, and others.  It is perhaps worthy at this stage to explicitly declare  that we are talking of a {\it generalization} of the BG theory, {\it by no means of an alternative to it}.

\section{LEARNING FROM BOLTZMANN-GIBBS STATISTICAL MECHANICS}

\begin{center}
{\bf A. Differential equation}
\end{center}

Which is the simplest ordinary differential equation? No doubt it is
\begin{equation}
\frac{dy}{dx} = 0 \ ,
\end{equation}
whose solution (with $y(0)=1$) is $y=1$. What could be considered as the second in simplicity? Surely
\begin{equation}
\frac{dy}{dx}=1\ ,
\end{equation}
whose solution is $y=1+x$. And the next one? It is tempting to say
\begin{equation}
\frac{dy}{dx}=y \ ,
\end{equation}
whose solution is $y=e^x$. Its inverse is $y=\ln x$, which coincides, by the way, with the celebrated Boltzmann formula (called {\it Boltzmann principle} by Einstein)
\vspace*{1cm}
\begin{equation}
S_{BG}=k\ln W \ ,
\end{equation}
where $k$ is Boltzmann constant, and $W$ is the measure of the space where the system is allowed to ``live'', taking into account total energy and similar constraints. So, typically, if we have an isolated $N$-body Hamiltonian system (microcanonical ensemble in Gibbs notation), W is the dimensionless Euclidean 
{\it measure} (i.e., (hyper)volume) of the fixed-energy Riemann (hyper)surface in phase space (Gibbs'
$\Gamma$-space) if the system microscopically follows {\it classical dynamics}, and it is the {\it dimension} of the associated Hilbert space if the system microscopically follows 
{\it quantum dynamics}. In what follows we indistinctively refer to classical or quantum systems. We shall nevertheless use, for simplicity, the wording ``phase space'' although we shall write down formulas where W is a natural number.

If we introduce a natural scaling for $x$ (i.e., if $x$ carries physical dimensions) we must consider, instead of Eq. (3),
\begin{equation}
\frac{dy}{dx}=ay\ ,
\end{equation}
in such a way that $ax$ is a dimensionless variable. The solution is now
\begin{equation}
y=e^{ax}\ .
\end{equation}
This differential equation and its solution appear to admit at least three physical interpretations that are crucial in BG statistical mechanics. Let us state them {\it without proof}. The {\it first} one is $(x,y,a)\rightarrow (t,\xi, \lambda )$, hence
\begin{equation}
\xi = e^{\lambda t}\ ,
\end{equation}
where $t$ is time, 
$\xi \equiv \lim_{\Delta X(0)\rightarrow 0}\frac{\Delta X(t)}{\Delta X(0)}$  is the 
{\it sensitivity to initial conditions}, and $\lambda$ is the (maximal) Lyapunov exponent associated with a typical phase-space variable
$X$ (the dynamically most unstable one, in fact). This
sensitivity to initial conditions (with $\lambda > 0$) is of course the cause of the mixing in phase space which will guarantee {\it ergodicity}, the well known dynamical justification for the entropy (4).

The {\it second} physical interpretation is given by 
$(x,y,a)\rightarrow (t,\Omega ,-1/\tau )$, hence
\begin{equation}
\Omega = e^{-t/\tau}\ ,
\end{equation}
where 
$\Omega \equiv \frac{{\cal O}(t)-{\cal O}(\infty )}{{\cal O}(0)-{\cal O}(\infty )}$, and 
$\tau$ is the characteristic 
time associated with the {\it relaxation} of a typical macroscopic observable ${\cal O}$ towards its value at the possible stationary state ({\it thermal equilibrium} for BG statistical mechanics). This relaxation occurs precisely because of the
sensitivity to initial conditions, which guarantees strong chaos (essentially Boltzmann's 1872 {\it molecular chaos hypothesis}). It was apparently Krylov the first to realize \cite{krylov}, over half a century ago, this deep connection. Indeed, $\tau$ typically scales like $1/\lambda$.

The {\it third} physical interpretation is given by $(x,y,a)\rightarrow (E_i,Zp_i,-\beta)$, hence
\begin{equation}
p_i=\frac{e^{-\beta E_i}}{Z}\left(Z\equiv \sum^W_{j=1}e^{-\beta E_j}\right) \ ,
\end{equation}
where $E_i$ is the eigenvalue of the $i$-th quantum state of the Hamiltonian (with its associated boundary conditions), $p_i$ is the probability of occurrence of the $i$-th state when the system is at its {\it macroscopic stationary state} in equilibrium with a thermostat whose temperature is $T\equiv 1/k\beta$ (canonical ensemble in Gibbs notation). It is a remarkable fact that the {\it exponential} functional form of the distribution which optimizes the BG generic entropy
\begin{equation}
S_{BG}=- k\sum^W_{i=1}p_i\ln p_i\ ,
\end{equation}
with the constraints
\begin{equation}
\sum^W_{i=1}p_i=1\ ,
\end{equation}
and
\begin{equation}
\sum^W_{i=1}p_iE_i=U\quad (U\equiv internal \ energy),
\end{equation}
{\it precisely is the inverse functional form of the same entropy under the hypothesis of equal probabilities}, i.e., $p_i=1/W(\forall_i)$, hence the {\it logarithmic} Eq. (4). To the best of our knowledge, there is (yet) no clear generic mathematical linking for this fact, but it is nevertheless true. It might seem at first glance a quite bizarre thing to do that of connecting the standard BG exponential weight to the solution of a (linear) differential equation, in contrast with the familiar procedure consisting in extremizing an entropic functional (Eq. (10)) under appropriate constraints (Eqs. (11) and (12)). It might be helpful to remind to those readers who so think that it is precisely through a differential equation that Planck heuristically found the celebrated black-body radiation law in his October 1900 paper \cite{planck}, considered by many as the beginning of the path that led to quantum mechanics.

Let us conclude the present remarks by saying that, when we stress that Eqs. (7), (8) and (9) naturally co-emerge within BG statistical mechanics, we only refer to the generic (or more typical) situations, {\it not to all} the situations. It is known, for example, that relaxation occurs through a power-law function of time at any typical second-order phase transition, whereas the Boltzmann-Gibbs weight remains exponential.

\begin{center}
{\bf B. Mean value}
\end{center}

The BG entropy (10) can be rewritten as the following mean value:
\begin{equation}
S_{BG}=k\langle\ln \frac{1}{p_i}\rangle \ ,
\end{equation}
where $\langle \cdots \rangle \equiv \displaystyle{\sum^W_{i=1}}p_i (\cdots )$. The quantity 
$\ln (1/p_i)$ is some
times called {\it surprise} \cite{watanabe} or {\it unexpectedness} \cite{barlow}. 
We notice that the averaged quantity has the {\it same} functional form as that corresponding to the equal probability case Eq. (4), where $1/p_i$ plays the role of $W$.

\begin{center}
{\bf C. Entropy composition law for independent systems}
\end{center}

Let us consider systems $A$ and $B$ as probabilistically independent, i.e., such that 
$p^{A+B}_{ij}=p^A_ip^B_j\ (\forall (i,j))$. We can immediately prove that entropy (10) satisfies the following property
\begin{equation}
S_{BG}(A+B)=S_{BG}(A)+S_{BG}(B)\ ,
\end{equation}
referred from now on as {\it extensivity}. This property is sometimes referred to as {\it additivity}, reserving the word  {\it extensivity} for the infinitely many body systems; we will
for simplicity not make such distinction here.

The {\it linear} property (14) of course encompasses the fact that, since 
$W_{A+B}=W_AW_B$, whenever we have equal probabilities, the logarithmic form (4) is absolutely fitted. For example, if we have $N$ independent coins (or dices), it is 
$W=2^N$ (or $6^N$), hence $S_{BG}=Nk\ln 2$ (or $S_{BG}=Nk\ln 6$). If we have, as another example, a $d=3$ regular lattice with ferromagnetic Heisenberg interactions between first neighbors at very high temperature, it is $W \sim A\rho^N$ (with $A>0,\ \rho > 1$, and 
$N\rightarrow \infty$), hence $S_{BG}\sim Nk\ln \rho$. In all these cases, we have 
$S_{BG} \propto N$, which precisely fits the Clausius concept of thermodynamic entropy. We shall discuss in Section II what can we do when the ubiquitous (but not obligatory) behavior $W(N)\sim \rho^N$ (with $N>>1$) is drastically violated, e.g., when 
$W\propto N^{\gamma}$, with $\gamma  > 0$, which {\it also} appears to be ubiquitous in both natural and artificial systems.

\begin{center}
{\bf D. Concavity}
\end{center}

Let us consider two probability distributions 
$\{p_i\}$ and $\{p'_i\}$ for a given system ($i=1,2,\cdots , W$). We shall define an {\it intermediate} distribution as follows:
\begin{equation}
p''_i\equiv \mu p_i+(1-\mu )p'_i\quad (0<\mu <1)\ , 
\end{equation}
An entropic functional $S(\{p_i\})$ is said {\it concave} if and only if
\begin{equation}
S(\{p''_i\})\geq \mu S(\{p_i\})+(1-\mu )S(\{p'_i\}) \;(\forall \{p_i\}, \forall \{p_i^\prime \}, \forall \mu) \;.
\end{equation}
It can be easily shown that $S_{BG}$ is concave. This property generically yields, within 
$BG$ statistical mechanics, {\it thermodynamical stability}, i.e., stability of the system with regard to energetic perturbations. For the canonical ensemble this implies 
$\partial^2S_{BG}/\partial U^2\leq 0$ (i.e., positive specific heat). For the microcanonical ensemble, the situation can in fact be more complex (see, for instance, \cite{gross}). Also, it is this property which makes that if we put into thermal contact two systems which are 
initially at different temperatures (of the same sign, usually positive), thermal equilibrium occurs at a temperature which is necessarily {\it intermediate} between the two initial ones. 
And this occurs with a variation of the total entropy which is necessarily an {\it increase}. We therefore see how central the concavity of the entropy is for both the $0^{th}$ and the 
$2^{nd}$ principles of thermodynamics.

\begin{center}
{\bf E. Stability or experimental robustness}
\end{center}

Lesche \cite{lesche} addressed in 1982 a very basic (and, curiously enough, widely unknown) property. A necessary condition for a (positive) statistical functional 
$O(\{p_i\})$ to be a physical quantity is \cite{lesche} that, under arbitrary small variations of the probabilities $\{p_i\}$, its relative variation remains small. We then say that 
$O(\{p_i\})$ is {\it stable} \cite{lesche} (we shall hereafter also use the expression 
{\it experimentally robust}, in order to avoid confusion with the thermodynamic stability mentioned in the previous subsection). To be more precise, if we consider two probability sets $\{p_i\}$ and $\{p'_i\}$ associated with $W$ microstates, the measure of the size of the deformation can be defined as follows \cite{lesche}
\begin{equation}
||p-p'||=\sum^W_{i=1} |p_i-p'_i|\ ,
\end{equation}
The condition of stability or experimental robustness of $O(\{p_i\})$ is then given by
\begin{equation}
||p-p'||<\delta_{\varepsilon}\quad \Longrightarrow R \equiv \left|
\frac{O(\{p_i\})-O(\{p'_i\})}{O_{max}}\right|<\varepsilon \;,
\end{equation}
for any $\varepsilon > 0$, with $\delta_{\varepsilon} > 0$ being independent from 
$W$, $O_{max}$ being the maximal value that $O(\{pi\})$ can attain. This implies, 
in particular, that $\displaystyle{\lim_{\varepsilon \rightarrow 0}}\ 
\displaystyle{\lim_{W\rightarrow \infty}}\ R=
\displaystyle{\lim_{W\rightarrow \infty}}\ 
\displaystyle{\lim_{\varepsilon \rightarrow 0}}\ R=0$.

It can be shown \cite{lesche} that $S_{BG}$ (for which $O_{max}=k\ln W$) is experimentally robust.

\begin{center}
{\bf F. Finite entropy production per unit time}
\end{center}

Let us consider the allowed phase space (assumed continuous and finite) of our system, and let us partition it in 
$W>>1$ little cells. Assume next an ensemble of $M>>1$ initial conditions, all inside one of the $W$ little cells, and let the microscopic dynamics of the system to run. If the system is strongly chaotic (i.e., if it has positive Lyapunov exponents), the points will quickly spread everywhere inside the allowed region. We might define a set of probabilities 
$p_i\equiv M_i(t)/M \ (i=1,2,\cdots , W)$, where $M_i(t)$ is the number of points inside the $i$-th cell at time $t$ $\left(\displaystyle{\sum^W_{i=1}}M_i(t)=M,\ \forall t\right)$. The entropy 
$S_{BG}(t)$ will be zero at t = 0, and will start increasing as t goes on, finally saturating at some value (which is $k\ln W$ if the system is {\it ergodic}; more precisely, if it equally visits all regions of the allowed phase space). The quantity
\begin{equation}
K_{BG}\equiv \lim_{t\rightarrow \infty}\lim_{W\rightarrow \infty}\lim_{M\rightarrow \infty}
\frac{S_{BG}(\{p_i(t)\})/k}{t}
\end{equation}
will in general be {\it finite} and equal to the Kolmogorov-Sinai entropy (which constitutes a slightly different, more familiar in the mathematical community, definition of the same concept). This finiteness follows from the Pesin theorem, which can be rewritten essentially as follows
\begin{equation}
K_{BG}=\sum_r\lambda_r\ ,
\end{equation}
where $r$ runs over all the positive Lyapunov exponents of the system.

\section{USING WHAT WE LEARNT FROM BOLTZMANN-GIBBS STATISTICAL MECHANICS TO GENERALIZE IT}

There are several other properties than those discussed above, which also specifically characterize BG statistical mechanics, but we shall restrict the present analysis to only those, i.e., differential equations, mean value, entropy composition law, concavity, experimental robustness, and entropy production. For further --- and consistent --- characterizations, see \cite{tsallisgellmann}. As we already mentioned, there is of course no logical-deductive manner to generalize a physical theory. Or, to put it on more general grounds, there is no generic or unique way of generalizing a logically consistent set of axioms into another one which also is logically consistent and which, by construction, recovers the original one as a particular case. 
It is therefore only metaphorically that we shall use, in what follows, the mathematical structure of BG statistical mechanics in order to generalize it.

\begin{center}
{\bf A. Differential equations}
\end{center}

Can we unify Eqs. (1), (2) and (3) in a {\it single} differential equation? Yes, through
\begin{equation}
\frac{dy}{dx}=a+by\ .
\end{equation}
Can we do it {\it minimally}, with only one parameter? (instead of two, namely $a$ and $b$). Yes, {\it out of linearity}, through
\begin{equation}
\frac{dy}{dx}=y^q\quad (q\in {\cal R})
\end{equation}
Eqs. (1), (2) and (3) are respectively recovered for 
$q \rightarrow  -\infty$, $q=0$ and $q=1$. The solution of Eq. (22) 
(with $y(0)=1$) is given by
\begin{equation}
y=[1+(1-q)x]^{1/(1-q)} \equiv  e_q^x \;\;\; (e^x_1=e^x)\ .
\end{equation}
The inverse function of the {\it q-exponential} is the {\it q-logarithm}, defined as follows
\begin{equation}
y=\frac{x^{1-q}-1}{1-q}\equiv \ln_qx\quad (\ln_1x=\ln x)\ .
\end{equation}
It should be clear that {\it these arguments about differential equations have by no means a provatory nature nor intention}. They are given here to provide some specific ``feeling" about {\it linearity} and {\it nonlinearity}, thus providing some intuitive plausibility to the generalization that we propose in what follows. On these grounds, we may expect
the Boltzmann principle, Eq. (4), to be generalized, 
for equal probabilities, as follows
\begin{equation}
S_q(p_i=1/W,\ \forall i)=k\ln_q W=k\frac{W^{1-q}-1}{1-q}\ ,
\end{equation}

As for the BG case, if $x$ carries a physical dimension, we must consider, instead of Eq. (22),
\begin{equation}
\frac{dy}{dx}=a_qy^q\;\;\;(a_1=a)\ ,
\end{equation}
hence
\begin{equation}
y=e^{a_qx}_q\ ,
\end{equation}
As for the BG case, we expect this solution to admit at least three different physical interpretations. The first one corresponds to the sensitivity to initial conditions
\begin{equation}
\xi =e^{\lambda_qt}_q \ ,
\end{equation}
where $\lambda_q$ generalizes the Lyapunov exponent or coefficient. Expression (28) was conjectured in 1997 \cite{TPZ}, and, for unimodal maps, proved recently \cite{baldovinrobledo,baldovinrobledo2}.

The second interpretation corresponds to relaxation, i.e.,
\begin{equation}
\Omega =e^{-t/\tau_q}_q\ ,
\end{equation}
There is (yet) no proof of this property, but there are several verifications (see, for instance, \cite{weinstein} for a quantum chaotic system).

The third interpretation corresponds to the energy distribution at the stationary state, i.e.,
\begin{equation}
p_i=\frac{e^{-\beta_qE_i}_q}{Z_q}\ 
\left(Z_q\equiv \sum^W_{j=1}e^{-\beta_q E_j}_q\right)\ .
\end{equation}
This is precisely the form that comes out from the optimization of the generic entropy $S_q$ under appropriate constraints \cite{tsallis,tsallis2,tsallis3}. This form has been observed in a large variety of situations (see \cite{review} for mini-reviews).

Before closing this subsection, let us stress that there is no reason for the values of $q$ appearing in Eqs. (28), (29) and (30) be the same. Indeed, if we respectively denote
them by $q_{sen}$ ($sen$ stands for {\it sensitivity}), $q_{rel}$ ($rel$ stands for 
{\it relaxation}) and $q_{stat}$ ($stat$ stands for {\it stationary state}), we typically (but not necessarily) have that $q_{sen}\leq 1$, $q_{rel}\geq  1$ and 
$q_{stat} \geq  1$. The possible connections between all these entropic indices are not (yet) known in general. However, for the edge of chaos of the z-logistic maps (see \cite{lyratsallis,costa,moura,borges} and references therein) we do know some important properties. If we consider the multifractal 
$f(\alpha )$ function, the fractal or Hausdorff dimension $d_f$ corresponds to the maximal height of $f(\alpha )$; also, we may denote by $\alpha_{min}$ and $\alpha_{max}$ the values of $\alpha$ at which $f(\alpha )$ vanishes (with $\alpha_{min} < \alpha_{max}$), It has been proved \cite{baldovinrobledo,lyratsallis} that
\begin{equation}
\frac{1}{1-q_{sen}}=\frac{1}{\alpha_{mim}}-\frac{1}{\alpha_{max}}\ .
\end{equation}
Moreover, there is some numerical evidence \cite{moura} suggesting
\begin{equation}
\frac{1}{q_{rel}-1}\propto (1-d_f)\ .
\end{equation}
Unfortunately, we know not much about $q_{stat}$, but it would not be surprising if it was closely related to $q_{rel}$. They could even coincide, in fact.

\begin{center}
{\bf B. Mean value}
\end{center}

Since we have seen in the previous subsection that the logarithmic function naturally generalizes into the  $q$-logarithmic one, let us define
\begin{equation}
S_q=k\langle \ln_q\ \frac{1}{p_i}\rangle \ ,
\end{equation}
where we may call $\ln_q(1/p_i)$ the {\it q-surprise} or {\it q-unexpectedness}. Then, using Eq.(24), it is straightforward to obtain
\begin{equation}
S_q=k\ \frac{1-\displaystyle{\sum^W_{i=1}p_i^q}}{q-1}\quad (S_1=S_{BG})\ ,
\end{equation}
which is the entropy on which we shall base the present generalization of BG statistical mechanics \cite{tsallis,tsallis2,tsallis3}. This entropy of course recovers Eq. (25) for equal probabilities.

\begin{center}
{\bf C. Entropy composition law for independent systems}
\end{center}

If we consider now the same two probabilistically independent systems $A$ and $B$ that we assumed before, we straightforwardly obtain
\begin{equation}
\frac{S_q(A+B)}{k}=\frac{S_q(A)}{k}+\frac{S_q(A)}{k}+(1-q)
\frac{S_q(A)}{k}\ \frac{S_q(B)}{k}\ .
\end{equation}
We re-obtain Eq. (14) in the limit $(1-q)/k\rightarrow 0$. Since $S_q$ is always nonnegative, we
 see that, if $q<1$ $(q>1)$, we have that $S_q(A+B)>S_q(A)+S_q(B)   \; (S_q(A+B)<S_q(A)+S_q(B))$, which shall be referred as the {\it superextensive (subextensive)} case. It is from this property that the expression {\it nonextensive statistical mechanics} was coined.

Let us address now the case $W \propto N^{\gamma}(\gamma >0;\ N>>1)$ that we mentioned earlier. If we replace this into Eq. (25) we obtain $S_q\propto N^{\gamma (1-q)}$ if $q<1$. Consequently, it exists a special and unique value of $q$, namely $q^{\ast}=1-(1/\gamma )<1$, for which we have $S_{q^{\ast}}\propto N$, once again in agreement with the Clausius idea of entropy! This only occurs because the case $W \propto N^{\gamma}$ is incompatible with any hypothesis of probabilistic independence (even in the limit $N\rightarrow \infty$). Amusingly enough, we see that it might happen that the entropy $S_q$ with $q\neq 1$, which is {\it nonextensive} for independent systems, can be {\it extensive} when we consider a special class of systems which include strong dependence. More explicitly, if we asymptotically have $W\propto \rho^N$, it follows (for $N_A$ and $N_B$ large and independent) $W(A+B)\propto \rho^{N_A+N_B}
=\rho^{N_A}\rho^{N_B}\propto W(A)W(B)$, hence we must choose $q^{\ast}=1$ to have a finite limit for the entropy per particle. Whereas, if we asymptotically have 
$W\propto N^{\gamma}$, it follows that $W(A+B)\propto (N_A+N_B)^{\gamma}\neq
N^{\gamma}_AN^{\gamma}_B \propto W(A)W(B)$, and the way for still having a {\it finite} limit for the entropy per particle is choosing $q^{\ast}=1-(1/\gamma )$.

It is interesting to notice \cite{baldovinrobledo2} that if we replace $W \propto N^{\gamma}$ into 
$S_{2-q}$ we obtain $S_{2-q}\propto N^{\gamma (q-1)}$, hence 
$S_{2-q^{\ast \ast}}\propto N$ if $q^{\ast \ast}=1+(1/\gamma )>1$. Naturally 
$q^{\ast \ast}=2-q^{\ast}$.

In fact, this remark can be trivially generalized. Consider $\kappa (q)$, where $\kappa$ is any monotonic continuous function satisfying $\kappa (1)=1$. If we focus on the decreasing ones, an example of such function is the one already considered, namely 
$\kappa =2-q$. Another one is $\kappa =1/q$. Other possibilities can be obtained by successively applying the two just mentioned, i.e., $\kappa  =1/(2-1/q)=q/(2q-1)$, or, the other way around, $\kappa = 2-1/(2-q)=(3-2q)/(2-q)$. (In fact, the four possibilities we have just considered belong to the class $\kappa =(a+bq)/[(a+2b)q-b]$ with $(a,b)\in {\cal R}^2$, which is symmetric with regard to the axis $\kappa =q$). Then, for the case $W\propto  N^{\gamma}$ we are considering, we have that $S_{\kappa (q^{\ast})}\propto N$ with 
$q^{\ast}=\kappa^{-1}(1-(1/\gamma ))$, $\kappa^{-1}$ denoting the inverse function. If 
$\kappa (q)$ is an increasing (decreasing) function, we have that 
$q^{\ast}<1(q^{\ast}>1$). The considerations we have done in this paragraph might be not unrelated with the possible connections between the entropic indices $q_{rel}$, 
$q_{stat}$ and $q_{stat}$ mentioned earlier.

Finally, let us remind that Eq. (35) can be rewritten in the following extensive form \cite{tsallis}:
\begin{equation}
S^R_q(A+B)=S^R_q(A)+S^R_q(B)\ ,
\end{equation}
where
\begin{equation}
S^R_q\equiv \frac{\ln [1+(1-q)S_q/k]}{1-q}=
\frac{\ln \displaystyle{\sum^W_{i=1}}p^q_i}{1-q}
\end{equation}
is the Renyi entropy. This entropy is extremal at $1/W$ (equal probabilities), and attains the value $\ln W$ for {\it all} values of q. Clearly, this fact makes it useless for having a finite entropy per particle in the case 
$W \propto N^{\gamma}$. It is, nevertheless, an interesting functional for geometrically characterizing multifractals, as long known.

\begin{center}
{\bf D. Concavity}
\end{center}

It can be shown that $S_q(\{p_i\})$ is a concave (convex) functional for {\it all} positive (negative) values of $q$. It also follows that, for the canonical ensemble, the specific heat is necessarily positive for $q\geq 1$, but not necessarily for $0<q<1$.

Let us mention that Renyi entropy is concave for $q\leq 1$, but has no definite concavity (or convexity) for $q>1$. We consider this as a serious drawback for the use of Renyi entropy with 
$q>1$ for thermostatistical and thermodynamical purposes.

\begin{center}
{\bf E. Stability or experimental robustness}
\end{center}

Abe has recently proved \cite{abe} a remarkable fact, namely that $S_q$ is stable (experimentally robust) for {\it all} positive values of $q$. In contrast, Renyi entropy is unstable (experimentally fragile) for all values of $q\neq 1$ \cite{lesche}. Once again, we consider this a further serious drawback for the use of the Renyi functional as a basis for thermodynamics.

\begin{center}
{\bf F. Finite entropy production per unit time}
\end{center}

If the dynamics of the system is such that the Lyapunov exponents vanish, Eq. (20) 
provides $0=0$, which clearly is poorly informative. We would like to ``unfold''  this trivial equality, and know more about those zeros. The functional $S_q$ has been shown to enable precisely this in a variety of situations, including various low- and high-dimensional nonlinear dynamical systems, both conservative and dissipative, both discrete and continuous in time and space. The situation can be summarized as follows.

We generalize definition (19) as follows:
\begin{equation}
K_q\equiv \lim_{t\rightarrow \infty}\lim_{W\rightarrow \infty}\lim_{M\rightarrow \infty}
\frac{S_q(\{p_i(t)\})/k}{t}\quad (K_1=K_{BG})\ .
\end{equation}
What has been verified for vast classes of $K_{BG}=0$ nonlinear dynamical systems (see, for instance, \cite{LBRT}) is that a special and unique value of $q$ exists (coincident with 
$q_{sen}<1$ every time checking has been performed) such that $K_{q_{sen}}$ is finite, whereas 
$K_q$ vanishes (diverges) for all values of $q>q_{sen}$ $(q<q_{sen}$). In other words, for such systems, the entropy whose production per unit time is {\it finite} is $S_{q_{sen}}$, {\it not $S_{BG}$}.

The full characterization of the systems so behaving, and the full comprehension of their dynamical details in what concerns wandering in phase space, are still lacking. The scenario seems nevertheless relatively clear. If a chaotic sea (frequently single connected) exists in phase space, and we perform an average over many initial conditions (i.e., many choices for the little cell, among the $W$ cells existing in the partition, where we initially place the $M$ points) all over the entire allowed phase space, then the $q_{sen}=1$ entropy production is finite, {\it independently on whether the measure of occupancy is uniform} (i.e., ergodic in the entire allowed phase space) or not. But if the evolution in phase space is such that the system remains sensibly long times in regions where the structure is multifractal-like (possibly scale-free-like, in the sense of Barabasi et al, see \cite{barabasi} and references therein), then one expects $q_{sen}<1$. More details will be shown later in this paper.

The generalization of the Pesin theorem (Eq. (20)) along the present lines was conjectured in 1997 \cite{TPZ}, and has been recently proved by Baldovin and Robledo \cite{baldovinrobledo2} for unimodal one-dimensional maps. More precisely, they have proved that
\begin{equation}
K_{q_{sen}}=\lambda_{q_{sen}}
\end{equation}
We believe this to be one of the cornerstones of the entire theory.

It is clear that the comments we did earlier concerning the behavior of $S_q$ as a function of 
$N$ are applicable here as a function of $t$. In other words, the entropy production per unit time is finite also for $S_{\kappa (q)}(t)$, i.e., if $K_{q_{sen}}$ is finite, the same occurs for $K_{\kappa (q_{sen})}$. Also, it is clear from Eq. (37) that for all those $q\neq  1$ systems for which $S_q(t)$ is asymptotically linear, Renyi entropy is {\it not}. Therefore, for such systems, there is no value of $q$ (excepting of course whenever $q_{sen}=1$) for which 
$S^R_q$ asymptotically could yield a finite entropy production per unit time. 

\section{THE CANONICAL ENSEMBLE AND CONNECTION TO THERMODYNAMICS}

In the previous Sections we have addressed the question of {\it which one}, among the infinite possible generalizations of $S_{BG}$, we want to {\it postulate} as the basis for generalizing BG statistical mechanics. We have decided it will be $S_q$. Now, if we have an entropic form, in principle {\it any} entropic form, and are interested in cybernetics, control theory, information theory, and related matters, there are many things that we can do just with that. If we are, however, physicists, we might naturally think of doing statistical mechanics and thermodynamics. In other words, we will consider the {\it energy} (which represents the physical support of the system) in addition to the {\it entropy} (which represents our information about that physical support). To do so, we shall first address what we consider to be the ``Sancta Sanctorum" of statistical mechanics: conservative Hamiltonian systems, i.e., the systems Boltzmann and Gibbs themselves had in mind. More specifically, we shall address a large system in contact with an even (much) larger thermostat ({\it canonical ensemble}).   
Following along Gibbs' variational path, we shall extremize $S_q$ with the norm constraint (11), and with a supplementary constraint related to the Hamiltonian ${\cal H}$ of the system, namely \cite{tsallis3}
\begin{equation}
\langle  {\cal H} \rangle_q \equiv   \frac{\sum_{i=1}^W p_i^q E_i}{\sum_{j=1}^W p_j^q } =U_q \;,
\end{equation}
with $  \langle ... \rangle_1 = \langle ... \rangle$; $P_i \equiv p_i^q /\sum_{j=1}^W p_j^q$ is referred to as the {\it escort distribution} \cite{beckschloegl}, and $\{E_i\}$ is the set of eigenvalues associated with ${\cal H}$ and the corresponding boundary conditions. We discuss below the reasons which make desirable the use of $P_i$ instead of just $p_i$ for defining the energy constraint.
The extremizing distribution corresponds to (meta)equilibrium and is straightforwardly shown to be given by
\begin{equation}
p_i= \frac{ e_q^{-\beta_q(E_i-U_q)}     }{{\bar Z}_q} \;,
\end{equation}
with
\begin{equation}
{\bar Z}_q \equiv \sum_{j=1}^W e_q^{-\beta_q(E_j-U_q)} \;,
\end{equation}
and
\begin{equation}
\beta_q \equiv \frac{\beta}{\sum_{j=1}^Wp_j^q} \;,
\end{equation}
$\beta \equiv 1/kT$  being the Lagrange parameter associated with constraint (40).
We easily verify that $q=1$ recovers the standard BG weight, $q>1$ implies in a {\it power-law tail} at high values of $E_i$, and $q<1$ implies in a {\it cutoff} at high values of $E_i$.  The (meta)equilibrium distribution (41) can be rewritten as follows: 
\begin{equation}
p_i= \frac{ e_q^{-\beta_q^\prime E_i}     }{Z_q^\prime} \;,
\end{equation}
with
\begin{equation}
Z_q^\prime \equiv \sum_{j=1}^W e_q^{-\beta_q^\prime E_j} \;,
\end{equation}
and
\begin{equation}
\beta_q^\prime \equiv \frac{\beta_q}{1+(1-q) \beta_qU_q}\;.
\end{equation}
This form is particularly convenient for many applications where comparison with experimental or computational data is involved. 

From the preceding results the connection to thermodynamics can be derived. In fact, the entire Legendre transformation structure of thermodynamics is $q$-invariant. In particular, it can be proved that
\begin{eqnarray}
\frac{1}{T} \equiv k{\beta}=\frac{\partial S_q}{\partial U_q}\;, \nonumber
\end{eqnarray}
as well as
\begin{eqnarray}
F_q \equiv U_q-\frac{S_q}{k{\beta}}= -\frac{1}{\beta} \ln_q Z_q\;, \nonumber
\end{eqnarray}
where
\begin{eqnarray}
\ln_q Z_q = \ln_q {\bar Z}_q - \beta U_q\;.  \nonumber
\end{eqnarray}
Also, it can be proved that
\begin{eqnarray}
U_q =-\frac{\partial}{\partial \beta} \ln_q Z_q \;, \nonumber
\end{eqnarray}
as well as
\begin{eqnarray}
C_q \equiv T \frac{\partial S_q}{\partial T}    =\frac{\partial U_q}{\partial T} = -T \frac{\partial^2F_q}{\partial T^2}      \;.\nonumber
\end{eqnarray}

The form adopted for constraint (40) (instead of the usual Eq. (12)) is at first sight astonishing, and surely demands clarification. It satisfies a remarkable set of (intertwined) properties, which we list now.

(i) It satisfies, exactly as it happens when using $p_i$, the basic property that the mean value of a constant is the same constant. This was not so in the version developed in 1991 \cite{tsallis2}. Chronologically speaking, to satisfy this property only became compelling when it became clear that the $q$-generalization concerned a problem of lack of ergodicity, and not any unusual norm-preservation anomaly. 

(ii) It makes that the addition of {\it microscopic} energies of independent systems preserves, at the {\it macroscopic} level, {\it exactly the same form}. In other words,  $E_{ij}^{A+B}=E_i^A + E_j^B$ with $p_{ij}^{A+B}=p_i^A p_j^B$ implies $U_q^{A+B}=U_q^A+U_q^B$. This property surely has an important role for the present generalization to satisfy the first principle of thermodynamics. This property was not satisfied in the 1991 version \cite{tsallis2}.  

(iii) The distribution (41) obtained for the stationary state is invariant under change of the zero of energies, i.e., under uniform translation of the microscopic energies. In more specific terms, if we add $E_0$ to all energies $E_i$, the same $E_0$ is, through Eq. (40), added to $U_q$, hence $E_i-U_q$ does not change, {\it nor does $p_i$}. This property was not explicitly satisfied in the early 1988 and 1991 versions \cite{tsallis,tsallis2}.

(iv) Abe and Rajagopal showed \cite{aberajagopal} that the traditional steepest descent method  (long ago used by Darwin and Fowler to discuss BG statistics) naturally leads to the escort distribution in the energy constraint.  It is basically related to the simple property $d e_q^x/dx = (e_q^x)^q$.

(v) The form of constraint (40) makes that, under the optimization of $S_q$, the Lagrange $\alpha$-parameter (the one associated with the norm constraint) appears in a function which naturally {\it factorizes} out of the sum over all states $\{i\}$. This property warranties the definition of a partition function which, as usually, depends on the Lagrange $\beta$-parameter but {\it not on $\alpha$}. This interesting mathematical fact was present in the 1991 version, but not in the version developed in 1988.

(vi) If we express the two $q>1$ canonical-ensemble constraints in the space of the energy, for high energy they asymptotically behave as indicated now:
\begin{eqnarray}
\sum_i p_i &\simeq&  \int_{constant}^\infty dE \,g(E)p(E) \nonumber \\
&\propto& \int_{constant}^\infty dE\,g(E) /E^{1/(q-1)} \nonumber
\end{eqnarray}
and
\begin{eqnarray}
\sum_ip_i^q E_i / \sum_j p_j^q &\propto& \int_{constant}^\infty dE \,g(E) E/E^{q/(q-1)} \nonumber \\
&=& \int_{constant}^\infty dE\,g(E) / E^{1/(q-1)} \nonumber
\end{eqnarray}
where $g(E)$ is the density of states.  We can observe that the domain of $q$ where the two constraints  are {\it finite} (hence mathematically and experimentally well defined) is the {\it same}; such nice property does {\it not} occur if we define the energy constraint using $p_i$ instead of the escort distribution. For example, if we assume the simple case where the high energy approximation is given by $g(E) \sim E^\delta$ ($\delta\in\Re$), the convergence of {\it both} integrals is guaranteed for $ 1/(q-1)-\delta >1 $, hence $q<(2+\delta)/(1+\delta)$. This feature constitutes an ingredient of consistency within the theory, which was not present in the 1988 version.
Physically speaking, the internal energy $U_q$ can be seen as a measure, at a given temperature, of the {\it width} or {\it spread} of the distribution $p_i$ above the lowest possible energy. If we show to a practically-minded scientist say an exponential distribution [$p(x) \propto e^{-ax}$ with $a>0$ for $x\ge 0$, and $p(x)=0$ for $x<0$] and a power-law  distribution [$(p(x) \propto 1/(1+bx)$ with $b>0$ for $x \ge 0$, and $p(x)=0$ for $x<0$], and ask him (her) what roughly are the widths, he (she) will promptly check the width at about half value of the maximum for {\it both} cases, quite independently from the fact that the first moment of the first example is {\it finite}, whereas it {\it diverges} for the second one. The expectation value of the energy calculated with the escort distribution $P_i$, and {\it not} with $p_i$, precisely is a measure for such width, for {\it all cases}, independently from what happens for the first moment of the distribution. In other words, this is a {\it robust} manner for characterizing this particular constraint.

(vii) Let us analyze the same property as in (vi) but for a case where $\langle x \rangle=0$, and consequently we need to refer to the second moment $\langle x^2\rangle$.
The already mentioned L\'evy-like superdiffusion illustrates such a situation. If we show to the same practically-minded scientist as before say a Gaussian [$p(x) \propto e^{-ax^2}$] and a Lorentzian [$p(x) \propto 1/(1+bx^2)$] distribution, and once again ask him(her) what roughly are the widths, he (she) will once again  check the width at about half value of the maximum for {\it both} cases, quite independently from the fact that the second moment of the first one is {\it finite}, whereas it {\it diverges} for the second one. 
Fixing, for the present L\'evy-like example, $\Bigl[\int_{-\infty}^\infty dx\; x^2 [p_q(x)]^q\Bigr]/\Bigl[\int_{-\infty}^\infty dx\; [p_q(x)]^q\Bigr]$ precisely characterizes the width for {\it all} values of $q$ below 3 (which simultaneously is the upper bound for normalizability!) \cite{pratowidth,bukman,celialangevin}. The Lorentzian example just evoked corresponds to $q=2$. 

All the above (i-vii) points were satisfactorily settled in the 1998 version of nonextensive statistical mechanics \cite{tsallis3}, which is the one that we use since then, hence in the present review. For clarity on the situation, see the Table. However, in spite of the remarkable mathematical consistency of the 1998 version that one might check in this Table, the ultimate geometrical-dynamical  understanding  of escort distributions is still lacking and is object of current studies (see, for instance, \cite{abeescort}). The situation seems to have some analogy with the following {\it  gedanken} problem. Let us consider a $10\; cm$ linear object, and let us implement on it the construction of the well known Cantor set, where, at each hierarchy, the central third is deleted. The final object is a fractal with zero one-dimensional Lebesgue measure, and whose Hausdorff dimension is $d_f=\ln2 / \ln3$. What is the measure of such an object? It is clearly given by $(10\;cm)^{d_f}=10^{\ln 2/\ln3}\;cm^{\ln 2/\ln3} \simeq 4.27\;cm^{0.63}$. In other words, we ``start" with the entire Lebesgue measure $10\;cm$, but then we ``correct"! The function $\ln_q W \equiv (W^{1-q}-1)/(1-q)$ does something quite analogous: it ``"starts with the entire measure $W$ of the Gibbs $\Gamma$ space, and then it ``corrects". This would correspond to the fact that, for nonextensive systems, the entire phase space is dynamically {\it not} nearly entirely occupied (i.e., the system is {\it not} ergodic), but only a scale-free-like part of it (which depends on the initial conditions) is visited. Consistently, one expects some type of ``correction". It is not impossible that the escort distribution would precisely provide the necessary ``correction". It is perhaps not unworthy to warn the reader that this is but a possible scenario, which remains to be proved.

\section{CALCULATING THE INDEX $\mathbf{\lowercase{q}}$ A PRIORI}

For the present theory to be complete, it is obviously necessary to be able to calculate $q$ a {\it priori}, i.e., from first principles. Consistently with the thought of Einstein \cite{einstein}, Cohen \cite{cohen}, Baranger \cite{baranger}, and many others, such a basic and crucial task can be accomplished nowhere else than in the analysis of the {\it microscopic} dynamics (classical, quantum, or others) of the system. However, very helpful connections can also be established at the level of
Langevin-like, Fokker-Planck-like, and similar equations, i.e., at the level of {\it mesoscopic} dynamics. In many occasions, natural and artificial systems whose microscopic or mesoscopic dynamics are either unknown or extremely complex have been addressed. For such systems, $q$ has been obtained through the fitting of experimental data. We are not addressing these cases here, but more details on several such examples can be found in \cite{review}. Our purpose here is to briefly mention several microscopic and mesoscopic determinations of $q$ that are available in the literature.

\begin{center}
{\bf A. Determination of $q$ from microscopic dynamics}
\end{center}

{\it (i) Low-dimensional dissipative systems (one- and two-dimensional dissipative maps)}\\

Let us consider the following one-dimensional dissipative maps (\cite{lyratsallis,costa,ugurcercle} and references therein):
\begin{equation}
x_{t+1} = 1-a |x_t|^z \;\;\;(z \ge 1) \;,
\end{equation}
referred to as the {\it $z$-logistic map}, 
and
\begin{equation}
x_{t+1} = d \,cos (\pi |x_t-1/2|^{z/2}) \;\;\;(z \ge 1) \;.
\end{equation}
referred to as the {\it $z$-periodic map}. They belong to the same Feigenbaum-Coullet-Tresser universality classes. More specifically, they share at their edge of chaos (e.g., for $z=2$,  $a_c=1.4011...$, and $d_c=0.8655...$), the same Feigenbaum universal constants $\delta_F(z)$ and $\alpha_F(z)$. 

Let us also consider (\cite{ugurcercle} and references therein) a family of one-dimensional dissipative maps which belong to universality classes that are {\it different} from the just mentioned: 
\begin{equation}
\theta_{t+1}= \Omega + [\theta_t-(1/2\pi) sin (2 \pi \theta_t)]^{z/3}  \;\;\;(z>0; \;0<\Omega<1) \;.
\end{equation}
referred to as the {\it $z$-circle map}. The edge of chaos occurs, for $z=3$ (the usual case), at $\Omega_c=0.6066...$. For all three maps (47), (48) and (49), the following relations have been verified  
\begin{equation}
\frac{1}{1-q_{sen}(z)}= \frac{1}{\alpha_{min}(z)} - \frac{1}{\alpha_{max}(z)} =\frac{(z-1) \ln \alpha_F(z)}{\ln b} \,
\end{equation}
where, as mentioned earlier, $\alpha_{min}$ ($\alpha_{max}$) is the minimal (maximal) value of $\alpha$ in the  $f(\alpha)$ multifractal function ($f(\alpha_{min})=f(\alpha_{max})=0$), and where $b=2$ for maps (47) and (48) and $b= (\sqrt 5 +1)/2 = 1.6180...$ ({\it golden mean})     for maps (49).

It is clear that the dissipative Henon map
\begin{eqnarray}
x_{t+1}&=& 1-a |x_t|^z +y_t \\ \nonumber
y_{t+1}&=& bx_t\;\;\;\;\;\;(0\le |b|<1)
\end{eqnarray}
belongs to the same universality class as the logistic map. Checking this case is not really necessary; nevertheless, the numerics have also been directly performed \cite{tirnaklihenon}. Consistency with the logistic map has indeed been verified.  

To be more explicit, $q_{sen}(z)$ has been calculated through various independent procedures, namely the sensitivity to the initial conditions \cite{TPZ,baldovinrobledo}, the multifractal function $f(\alpha)$ \cite{lyratsallis}, and the finiteness of the entropy production per unit time \cite{baldovinrobledo2,LBRT}. A fourth method has been used as well, which we describe now. By measuring the shrinking of the Lebesgue measure, Moura et al obtained $q_{rel}(z)$ for the $z$-logistic maps. For example $q_{rel}(2) \simeq 2.41$. A connection has been recently obtained \cite{borges} between $q_{sen}$ and $q_{rel}$, namely
\begin{equation}
q_{rel}(z,\infty) - q_{rel}(z,W) \propto 1/W^{|q_{sen}(z)|} \,
\end{equation}
where $W$ is the number of cells in which the phase space has been partitioned (the larger $W$, the thinner the graining). This equation can be used to simultaneously determine $q_{sen}(z) $ and $q_{rel}(z)= q_{rel}(z,\infty)$. 

To illustrate all these methods, let us mention that, since $\alpha_F(2)$ is known with at least 1018 digits for the logistic map, the corresponding value for 
\begin{eqnarray}
q_{sen}(2) =  0.2445... \nonumber
\end{eqnarray} 
is known with the same number of digits. Cases like this one completely disqualify the critical --- and, as we see, unfounded --- remark that we hear occasionally, namely that $q$ is no more than a ``fitting parameter". \\

{\it (ii) Low-dimensional conservative systems (two- and four-dimensional conservative maps)}\\

The standard map is defined as follows:
\begin{eqnarray}
\theta(t+1)&=&p(t)+\frac{a}{2\pi}\sin\,[2\pi \theta(t)]+\theta(t)~~~\rm{(mod\;1)},
\label{standard}\\
p(t+1)&=&p(t)+\frac{a}{2\pi}\sin\,[2\pi \theta(t)]~~~~~~~~~~~~\rm{(mod\;1)}\nonumber
\end{eqnarray}
$(a\in{\mathbb R},\;\;t=0,1,...)$;
$2\pi p$ may in fact be regarded as the angular
momentum of a free rotor subject to angle-dependent impulses of strength
$a$ at unit intervals of time.

This two-dimensional map is conservative and simplectic. Also, it is symmetric with regard to $p=1/2$. For $|a|>>1$, the system is very chaotic (large Lyapunov exponent) and ergodic. When $a$ approaches $a_c=0.9716...$ from above, the region around $p=1/2$ gradually becomes isolated from the rest of the phase space. Therefore, if we define a kind of ``dynamical temperature" $T(t)\equiv  \langle (p(t))^2 \rangle - \langle p(t) \rangle^2 $, a nonzero-measure region exists in phase space where, if we start inside it at $t=0$, we observe two plateaux in $T(t)$ \cite{brigatti}. The duration of the first plateau diverges when $a \to a_c$. In other words, $\lim_{t \to \infty} \lim_{a \to a_c} T(a,t) \ne \lim_{a \to a_c}\lim_{t \to \infty}  T(a,t)$. The existence of such a nonuniform convergence is, as we shall see later on, a distinctive feature which already suggests the possible existence of some degree of nonextensivity. Indeed, anomalies are observed in both the sensitivity to the initial conditions and in the entropy production per unit time. These anomalies appear to be consistent with each other, and they enable the characterization of an index $q_{sen}<1$.  However, this index is only an effective one,  and depends on $a$. For example, for $a=0.6$, we can verify that $q_{sen} \simeq 0.3$, but a permanent drift of $q_{sen}$ is observed towards zero, when we average over many initial conditions all over the entire phase space and consider values of $a$ gradually approaching zero.  The limit $a=0$ corresponds of course to the integrable case of the map. Details can be found in \cite{ananos}.

The two-dimensional map has no Arnold diffusion. To verify the influence of dimensionality we also considered two standard maps coupled  as follows:
\begin{eqnarray}
p_1(t+1)&=&p_1(t)+\frac{a_1}{2\pi}\sin\,[2\pi \theta_1(t)],
\label{standard_standard}\\
p_2(t+1)&=&p_2(t)+\frac{a_2}{2\pi}\sin\,[2\pi \theta_2(t)],\nonumber\\
\theta_1(t+1)&=&\theta_1(t)+p_1(t+1)+b\;p_2(t+1),\nonumber\\
\theta_2(t+1)&=&\theta_2(t)+p_2(t+1)+b\;p_1(t+1),\nonumber\\
\nonumber
\end{eqnarray}
where $a_1,a_2,b\in{\mathbb R},\;t=0,1,...$,  and all variables are defined 
$\rm{mod\;1}$. If the coupling constant $b$ vanishes the two standard maps decouple;
if $b=2$ the
points $(\theta_1,p_1,\theta_2,p_2)=(0,1/2,0,1/2)$ and $(1/2,1/2,1/2,1/2)$ are a $2$-cycle for all
$(a_1, a_2)$, hence we preserve, in phase space, the same referential  that we had for a single 
standard map. For a generic value of $b$, all relevant present results remain qualitatively the same.           
Also, we set $a_1=a_2\equiv\tilde a$ so that the system is invariant under
permutation $1\leftrightarrow 2$. 
This four-dimensional map is conservative and simplectic, and has Arnold diffusion. It exhibits two plateaus for $T(t)$, like the two-dimensional case (with $\tilde a_c=0$). The dependence of $q_{sen}(\tilde a)$ is not very different from the two-dimensional case. We are presently studying $N>>1$ coupled maps to verify the effects of the thermodynamic limit. This might or might not stabilize a nontrivial value for $q_{sen}$. 

Finally, let us mention in this section a connection with quantum chaos, namely for the quantum kicked top Hamiltonian. If we study the time evolution of the fidelity function (scalar product of the wave function of the original Hamiltonian with the wave function of a slightly perturbed Hamiltonian), three types of behaviors are typically observed. (i) In the regions corresponding to regular motion, the fidelity remains roughly constant; (ii) in the regions corresponding to standard chaos (with positive Lyapunov exponents for the associated classical system), the fidelity decreases exponentially with time (before entering the quantum interference region); (iii) at the border of the two regions, a power-law behavior has been recently observed \cite{weinstein} which precisely fits a $q$-exponential function. The value of $q_{rel}$ (associated with the zero perturbation limit) depends on the total momentum value $J$ of the top. It decreases with increasing $J$, and, in the limit $J \to\infty$ (classical limit), $q_{rel} \simeq 2.4$ can be extrapolated \cite{weinstein}. Further analysis is needed and welcome.\\

{\it (iii) High-dimensional dissipative systems (many-body dissipative models)}\\

A specific lattice Lotka-Volterra three-component model has been recently analyzed \cite{provata} in terms of entropy production per unit time. For $d$-dimensional growth, the following result has been obtained 
\begin{equation}
q_{sen}=1-1/d \;.
\end{equation} 
Details for the $d=1,2$ cases can be found in \cite{provata}; the $d=3,4$ cases have been studied by Anteneodo \cite{anteneodoprivate}. In fact the microscopic rules of the model are such that the growing droplet has a linear size which grows linearly with time $t$. The volume grows then as $t^d$. If we apply to this quantity the $q$-logarithmic function, we obtain that $S_q$ scales like $t$ if $(1-q)d=1$. From this observation, the result $q_{sen}=1-1/d$ follows immediately \cite{anteneodoprivate}. The possible connection with fractality as $d$ varies is yet unclear \cite{anteneodoprivate}.

Boltzmann $d$-dimensional Bravais lattice models for the Navier-Stokes equations for incompressible fluids have been recently revisited \cite{boghosian}. The imposition of the (physically desirable) Galilean invariance of the equations mandates an unique entropy, and that entropy precisely is $S_q$, with $q$ depending on some details of the model ($q$ is in fact analytically determined by a  transcendental equation). For example, for the single-mass single-speed model, it is 
\begin{equation}
q=1-2/d \;. 
\end{equation}

{\it (iv) High-dimensional conservative systems (many-body Hamiltonian systems)}\\

Hamiltonian systems is a central topic of statistical mechanics. We shall briefly review here some important results currently available for long-range-interacting classical systems. Let us first focus on the inertial $XY$ ferromagnetic model, characterized by the following Hamiltonian \cite{antoniruffo,celiaconstantino}:
\begin{equation}
{\cal H}=\sum_{i=1}^N  \frac{p_i^2}{2} +
  \sum_{i \ne j}  \frac{1-cos(\theta_i -\theta_j)}{r_{ij}^{\;\alpha}}~~~~(\alpha \ge 0),
\end{equation}
where $\theta_i$ is the $i-th$ angle and $p_i$ the 
conjugate variable   representing   the  angular momentum
(or the rotational velocity since, without loss of generality, unit moment of inertia is assumed).  
Notice that the summation in the potential is extended to all couples of spins (counted only once) 
and not restricted to first neighbors; for $d=1$, $r_{ij}=1, 2, 3,...$;  for $d=2$, $r_{ij}=1, \sqrt{2}, 2,...$; for $d=3$, $r_{ij}=1, \sqrt{2}, \sqrt{3}, 2, ...$. The first-neighbor coupling constant has been assumed, without loss of generality, to be equal to unity. This model is an inertial version of the well known $XY$ ferromagnet. Although it does not make any relevant difference, we shall assume periodic boundary conditions, the distance to be considered between a given pair of sites being the smallest one through the $2d$ possibilities introduced by the periodicity of the lattice. Notice that the two-body potential term has been written in such a way as to have zero energy for the global fundamental state (corresponding to $p_i=0$, $ \forall i$, and all $\theta_i$ equal among them, and equal to say zero). The $\alpha \to\infty$ limit corresponds to only first-neighbor interactions, whereas the $\alpha=0$ limit corresponds to infinite-range interactions (a typical Mean Field situation, frequently referred to as the HMF model \cite{antoniruffo}). 

In the limit $N \to \infty$, the quantity ${\tilde N} \equiv \sum_{i \ne j} r_{ij}^{\;-\alpha}$ converges to a finite value if $\alpha/d >1$, and diverges like $N^{1-\alpha/d}$ if $0 \le \alpha/d<1$ (like $\ln N$ for $\alpha/d=1$). In other words, the energy is extensive for $\alpha/d >1$ and nonextensive otherwise. In the extensive case (here  referred to as {\it short range interactions}; also referred to as {\it integrable interactions} in the literature), the thermal equilibrium (stationary state attained in the $t \to\infty$ limit) is known to be the BG one (see \cite{fisheretal}).  The situation is much more subtle in the nonextensive case ({\it long range interactions}). It is this situation that we focus on here. In order to conform to the most usual writing, we shall from now on replace the Hamiltonian ${\cal H}$ by the following rescaled one:
\begin{equation}
{\cal H^\prime}=\sum_{i=1}^N  \frac{p_i^2}{2} +
  \frac{1}{{\tilde N}}\sum_{i \ne j}  \frac{1-cos(\theta_i -\theta_j)}{r_{ij}^{\;\alpha}}~~~~(\alpha \ge 0),
\end{equation} 
The molecular dynamical results associated with this Hamiltonian (now artificially transformed into an extensive one for {\it all} values of $\alpha/d$) can be trivially transformed into those associated with Hamiltonian ${\cal H}$ by re-scaling time (see \cite{celiaconstantino}). 

Hamiltonian (58) exhibits in the microcanonical case (isolated system at fixed total energy $U$) a second order phase transition at $u \equiv U/N = 0.75$. It has anomalies both above and below this critical point. 

Above the critical point it has a Lyapunov spectrum which, in the $N \to \infty $ limit, approaches, for $0 \le \alpha/d \le1$, zero as $N^{-\kappa}$, where $\kappa(\alpha/d)$ decreases from $1/3$ to zero when $\alpha/d$ increases from zero to unity, and remains zero for $\alpha/d \ge 1$ \cite{celiaconstantino,giansanti}. It has a Maxwellian distribution of velocities \cite{vitoandreaconstantino}, and exhibits no aging \cite{aging}. Although it has no aging, the typical correlation functions depend on time as a $q$-exponential.  Diffusion is shown to be of the normal type.

Below the critical point (e.g., $u=0.69$), for a nonzero-measure class of initial conditions, a longstanding quasistationary (or metastable) state precedes the arrival to the BG thermal equilibrium state. The duration of this quasistationary state appears to diverge with $N$ like $\tilde N$ \cite{vitoandreaconstantino,cabral}. During this anomalous state, there is aging (the correlation functions being well reproduced by $q$-exponentials once again), and the velocity distribution is not Maxwellian, but rather approaches a $q$-exponential function (with a cutoff at high velocities, as expected for any microcanonical system). Anomalous superdiffusion is shown to exist in this state. The mean kinetic energy ($\propto T$, where $T$ is referred to as the dynamical temperature) slowly approaches the BG value from below, the relaxation function being once again a $q$-exponential one. During the anomalous aging state, the zeroth principle of thermodynamics and the basic laws of thermometry have been shown to hold as usual \cite{tsallisreply,moyano}. The fact that such basic principles are preserved constitutes a major feature, pointing towards the applicability of thermostatistical arguments and methods to this highly nontrivial quasistationary state. 

Although none of the above indications constitutes a proof that this long-range system obeys, in one way or another, nonextensive statistical mechanics, the set of so many consistent evidences may be considered as a very strong suggestion that so it is. Anyhow, work is in progress to verify closely this tempting possibility.

Similar observations are in progress for the Heisenberg version of the above Hamiltonian \cite{nobre}, as well as for a  $XY$ model including a local term which breaks the angular isotropy in such a way as to make the model to approach the Ising model  \cite{ernestoising}. 

Lennard-Jones small clusters (with $N$ up to 14) have been numerically studied recently \cite{doye}. The distributions of the number of local minima of the potential energy  with $k$ neighboring saddle-points in the configurational phase space can, although not mentioned in the original paper \cite{doye}, be quite well fitted with $q$-exponentials with $q=2$. No explanation is still available for this suggestive fact. Qualitatively speaking, however, the fact that we are talking of very {\it small} clusters makes that, despite the fact that the Lennard-Jones interaction is not a long-range one thermodynamically speaking (since $\alpha/d=6/3>1$), all the atoms sensibly see each other, therefore fulfilling a nonextensive scenario.  

Finally, as a last example of Hamiltonian systems where nonextensive features have been analytically obtained (or numerically observed) we would like to mention anomalous transport in an optical lattice. The distribution of velocities in such system has been recently shown \cite{lutz} to be, for a specific microscopic model, a $q$-distribution with
\begin{equation}
q=1 + \frac{44 E_R}{U_0}
\end{equation}
where  $E_R$ and $U_0$ are  microscopic parameters of the quantum optical problem.\\

{\it (v) Many-body growth models}\\

A growth model including {\it preferential} attachment  has been recently introduced by Albert and Barabasi \cite{barabasi}  as a prototype of emergence of the ubiquitous scale-free networks. At each time step, $m$ new links are added with probability $p$, or $m$ existing links are rewired with probability $r$, or a new node with $m$ links is added with probability $1-p-r$; all linkings are done with probability $\Pi(k_i)=(k_i+1)/\sum_j(k_j+1)$, where $k_i$ is the number of links of the $i-th$ node. The exact stationary state distribution of the number $k$ of links at each site can be written (although apparently not identified by the authors) as $p(k) \propto e_q^{-k/k_0}$ with
\begin{equation}
q=\frac{2m(2-r)+1 -p-r}{m(3-2r)+1-p-r}\;,
\end{equation}
$k_0>0$ being an explicit function of $(p,r,m)$.\\

{\it (vi) Diffusion in the diluted hypercube}\\

The diluted $N$-dimensional hypercube has been considered since long. Its percolation ``'threshold" is given \cite{erdos} by $p_c=\sigma +\frac{3}{2} \sigma^2 + \frac{15}{14} \sigma^3+...$ where $\sigma \equiv 1/(N-1)$. Strictly speaking, the quantity $p_c$ is an effective percolation threshold; indeed the exact percolation threshold clearly is $p_c=0$. Lemke and Almeida \cite{lemke} have recently studied the model where a random walker diffuses in this structure. More specifically they studied the entropy production for $N$ increasing up to 23, and a {\it finite} value was found only for $q=1$ at $p=1$ (full hypercube), and for $q=0.056$ at $p=p_c$. 

\begin{center}
{\bf B. Determination of $q$ from mesoscopic dynamics}
\end{center}

{\it (i) Nonlinear Fokker-Planck equations (Correlated anomalous diffusion)}\\  

In a variety of physical situations \cite{bukman}, it is appropriate to consider the following nonlinear Fokker-Planck-like equation (sometimes referred to as the {\it Porous Medium Equation}):
\begin{equation}
\frac{\partial p(x,t)}{\partial t}= -\frac{\partial}{\partial x}[F(x) p(x,t)] + D \frac{\partial^2 [p(x,t)]^\nu}{\partial x^2} \;\;\;(\nu \in {\cal R}) \;.
\end{equation}
If we assume that at $t=0$ we have the paradigmatic (and quite usual) distribution $p(x,0)=\delta(x)$, it can be shown \cite{bukman} that, for $F(x)=  k_1-k_2 x   $ ($k_1 \in {\cal R}$; $k_2 \ge 0$) and all $(x,t)$, the (stable) solution is given by
\begin{equation}
p(x,t) \propto e_q^{- \beta(t) [x-x_M(t)]^2} \;\;\;(q<3) \;,
\end{equation}
where $\beta(t)$ and $x_M(t)$ are smooth explicit functions of $t$, and 
\begin{equation}
q=2-\nu \;\;\;(q<3)\;. \\
\end{equation}

{\it (ii) Fractional-derivative Fokker-Planck equations (L\'evy anomalous diffusion)}\\ 

It is known  \cite{levy} that $p(x) \propto e_q^{-\beta x^2}$ ($q<3$) optimizes 
\begin{equation}
S_q =\frac{1-\int dx [p(x)]^q}{q-1}
\end{equation}
under appropriate constraints. If we convolute  $N$ times $p(x)$ ($N \to \infty$), we approach a Gaussian distribution if $q<5/3$ and a L\'evy $L_{\gamma_L}(x)$ one if $5/3<q<3$. L\'evy distributions are the solutions of the following equation $\frac{\partial p(x,t)}{\partial t}= D \frac{\partial^{\gamma_L} p(x,t)}{\partial x^{\gamma_L}} \;\;(0<\gamma_L <2)$.
The index $\gamma_L$ of these L\'evy distributions is related to $q$ as follows:
\begin{equation}
q=\frac{\gamma_L+3}{\gamma_L+1} \;\;\;(5/3<q<3) \;.
\end{equation}
In general, the L\'evy distribution and the $q$-Gaussian coincide only asymptotically (the tails). The unique exception is for $q=2 \gamma_L=2$, in which case they are identical. \\

{\it (iii) Nonlinear fractional-derivative  Fokker-Planck equations}\\

The equation 
\begin{equation}
\frac{\partial p(x,t)}{\partial t}= D \frac{\partial^\gamma [p(x,t)]^\nu}{\partial x^\gamma} \;\;\;(\nu \in {\cal R}; \; \gamma \in {\cal R}) \;.
\end{equation}
is an interesting one. The $\gamma =2$ particular instance reproduces the situation addressed with Eq. (61); the $\nu=1$ particular instance reproduces the partial differential equation above Eq. (65). We do not know the solution for the generic case. However, we do know it for the (new) case $\nu=(2-\gamma)/(1+\gamma)$ \cite{bologna}. It is given by
\begin{equation}
q=\frac{\gamma+3}{\gamma+1}= \frac{5+2 \nu}{3} \;
\end{equation}
with $0<\gamma, \nu<2$. As for the L\'evy case, it is only asymptotically that the solution coincides with the $q$-Gaussian function. More, and more complex, situations along these lines can be found in \cite{lenzimendestsallis}. \\

{\it (iv) Anomalous Langevin equations}\\

We may consider the following Langevin-like equation \cite{celialangevin}
\begin{equation}
{\dot x}= -\gamma x |x|^{2(s-1)} + x |x|^{s-1} \xi(t) + \eta(t)  \;\;\;\;(s>0),
\end{equation}
where $\xi(t)$ and $\eta(t)$ are independent and Gaussian-distributed zero-mean white noises, satisfying
\begin{equation}
\langle \xi(t) \xi(t^\prime) \rangle = 2 M \delta(t-t^\prime) \;\;\;\;(M \ge 0)
\end{equation}
and
\begin{equation}
\langle \eta(t) \eta(t^\prime) \rangle = 2 A \delta(t-t^\prime) \;\;\;\;(A >0)\;.
\end{equation}
$M$ and $A$ stand for {\it multiplicative} and {\it additive} respectively. If $\gamma \ge M(1-s)$, the distribution corresponding to the stationary state is given by
\begin{equation}
P(x) \propto e_q^{-\frac{(\gamma/s) +M}{2A} |x|^{2s}} \;,
\end{equation}
with
\begin{equation}
q=\frac{(\gamma/s) + 3M}{(\gamma/s) + M} \;.
\end{equation}
Notice that, interestingly enough, the case $s=M=1$ reproduces the results indicated in Eqs. (65) and (67).

Another interesting Langevin-like case is that of dichotomic colored noise addressed in \cite{caceres}. One may consider the following stochastic differential equation:
\begin{equation}
\frac{dV}{dt}= -\gamma V + \xi(t) \;,
\end{equation}
where $\xi(t)$ is a dichotomous noise of values $ \pm a$, satisfying 
\begin{equation}
\langle \xi(t + \tau) \xi \rangle =a^2 e^{-2\lambda |\tau|} \;.
\end{equation}
The stationary state probability is, although not written in this way in the original paper \cite{caceres}, given by $p(V) \propto e_q^{-\beta V^2}$ with
\begin{equation}
q= \frac{1-2\gamma/\lambda}{1-\gamma/\lambda}   \;\;\;\;\;\;(\gamma/\lambda \le1)\;, 
\end{equation}
$\beta$ being given by some smooth function of the model parameters.\\

{\it (v) Phenomenological approach of pressure fluctuations in multiphase flow}\\

A granular system has been experimentally studied \cite{gheorghiu} which consists of bubbling fluidized beds set in motion by a vertical gas stream. In addition to the experiments themselves, a phenomenological approach has been advanced. It provides, for the distribution of the pressure fluctuations, a $q$-Gaussian with
\begin{equation}
q=1+\frac{1}{\tau -1/2} \;\;\;(\tau > 3/2)\;,
\end{equation}
where $\tau$ is a phenomenological parameter of the bubble size distribution. We must warn the reader that the authors argue  \cite{gheorghiu} that, from their standpoint, no deep connection exists with nonextensive statistical mechanics. We do not necessarily agree with their interpretation. But it is not the aim of this short note to address this particular point in depth. We just refer the existence of this mathematical result in connection with those experiments. 

{\it (vi) Scalar granular gases}\\

Baldassarri et al \cite{baldassarri} have numerically and analytically discussed some isolated one-dimensional granular gases in a recent work.  A restitution coefficient $r \in [0,1)$ has been assumed for two-body collisions. The velocity distribution in the asymptotic cooling regime of a specific pseudo-Maxwell (Ulam's) model, has been shown to be a distribution, which, although the authors apparently have not identified it as such, precisely is a $q$-Gaussian with $q=3/2$. \\

\section{TOWARDS A GENERALIZED NONEXTENSIVE STATISTICAL MECHANICS}

A statistical mechanics can be seen as basically consisting in the appropriate choice of a physical entropic functional $S(\{pi\})$. Once this is done, we must consistently choose the constraints ($\sum^W_{i=1} p_i=1$, and possibly others) to be used for describing various physical situations (isolated, or in equilibrium with an infinitely large reservoir, or others). We must then extremize (typically maximize) the entropy under these constraints, in order to obtain the probability distribution corresponding to the stationary state, whenever it exists. Relaxation phenomena towards the distribution associated with the stationary state(s) can be characterized through $S(\{p_i(t)\})$, among others. The stationary state entropy is of course given by $S(\{p_i(\infty)\})$. The experimental information can provide (direct or indirect) evidence on $\{p_i(t)\}$, and on $S(t)=S\{p_i(t)\})$. But establishing the functional $S(\{p_i\})$ itself demands the knowledge of $S(t)$ for almost all possible trajectories. 
So, in practice, the functional $S(\{p_i\})$ is not deduced or measured, but found heuristically. This functional is in fact a mathematical construct which, for a variety of macroscopic physical quantities, can essentially replace (as the genius of Boltzmann showed to us) the detailed knowledge implied at the microscopic dynamical level.

The microscopic dynamics of an isolated system might be such that, for practically all possible initial conditions and as time increases, it quickly evolves towards an {\it uniform} occupation of phase space (either of the entire phase space, as for the $N\rightarrow \infty \;d=3$ short-range Heisenberg ferromagnet above its critical total internal energy, or of a simple and well defined part of it, as for the same ferromagnet {\it below} its critical internal energy, once a specific symmetry has been broken). The system is then said {\it ergodic}, and Boltzmann's basic equal-probability hypothesis indeed applies. The entropy to be used in such a case is widely known to be $S_{BG}$.

What happens, however, if the system has more complex dynamics, such that it tends, when isolated, to occupy phase space in a nontrivial nonuniform manner? This is where generalizations of BG statistical mechanics possibly become a natural way out. This imposes, in our opinion, the need of physical entropies {\it different} from the BG one.

The nonextensive entropy $S_q(\{p_i\})$

(i) satisfies $(\forall q > 0)$, among others, three mathematical properties (namely concavity, experimental robustness, and finiteness of the entropy production per unit time) that are 
{\it by no means trivial} to satisfy (e.g., Renyi entropy satisfies none of these 
$\forall q > 0)$,

(ii) yields a stationary distribution (the $q$-exponential function, which is asymptotically a power-law) which is ubiquitously found in natural and artificial systems, and

(iii) is in principle completely determined once its entropic index $q$ is calculated from microscopic or mesoscopic dynamics.

It is therefore allowed to consider $S_q$ as a quite strong candidate for physically extending the applicability of the methods of BG statistical mechanics to a specific class of
dynamical systems among those who do {\it not} comply with the usual ergodicity requirements. The rigorous characterization of this class is yet to be better understood, but we already dispose of a good basis for conjecturing that it concerns systems that mix slowly in phase space, yielding a long-standing (multi) fractal-like occupation of it (possibly with a 
{\it scale-free} structure, as it is nowadays called in the field of networks). Such is the case of various dissipative low-dimensional nonlinear dynamical systems at their edge of chaos, conservative low-dimensional non linear dynamical systems close to the frontier between
chaos and integrability, (possibly) long-range-interacting Hamiltonian systems, and others.

Are we compelled to stop here, or can we hope to similarly cover even more complex dynamics, eventually yielding stationary states (or sequences of quasistationary states) whose phase space structure is even more complex than the one just mentioned? The bottom
line is that we see no basic reason for stopping. However, a highly nontrivial problem must be solved before further progressing along this line, namely the proposal of
an appropriate entropy functional. There are at least two lines of thought (hereafter referred to as the {\it differential equation path}, and the {\it superstatistics path}) that converge
onto a solution of such a task. Let us next briefly mention both.

\begin{center}
{\bf A. The differential-equation path}
\end{center}

We have seen that BG and nonextensive statistical mechanics are deeply related to the differential equations (5) and (26) respectively. Can we unify them in such a way as to have a crossover between one and the other, depending on the value of the independent variable $x$? Yes, we can, and this has been advanced in 1999 \cite{bemski} for re-association in folded proteins, and recently used for a phenomenological theory for the flux of cosmic rays \cite{cosmic}. The associated differential equation we propose is as follows:
\begin{equation}
\frac{dy}{dx}=a_1y+(a_q-a_1)y^q\quad (y(0)=1)\ ,
\end{equation}
hence
\begin{equation}
y=\left[1-\frac{a_q}{a_1}\left(1-e^{(1-q)a_1x}\right)\right]^{\frac{1}{1-q}}
\end{equation}
Eq. (77) recovers Eq. (26) if we consider $a_1=0$. It recovers Eq. (5) if we consider 
$a_q=a_1$, and also if we just consider $q=1$. The existence of this last (convenient) possibility is the reason for which we presented the generalization in the form of Eq. (77), rather than in the totally equivalent form $\frac{dy}{dx}=a_1y+a'_qy^q$. An interesting situation occurs for the case $q<1$ with $0<a_1<<a_q$, as well as for the case $q>1$ with 
$a_q<<a_1<0$. For $0\leq x<<x_{crossover}\equiv [1/[(1-q)a_q]$ we have $y\sim e^{a_qx}_q$ (nonextensive statistical mechanics), whereas for $x>>x_{crossover}$ we have 
$y\sim e^{a_1x}$ (BG statistical mechanics).

We can further generalize Eq. (77) in a natural way, namely through
\begin{equation}
\frac{dy}{dx}=a_{q'}y^{q'}+(a_q-a_{q'})y^q\quad (y(0)=1)\ ,
\end{equation}
whose solution is a combination of hypergeometric functions (see \cite{bemski}). As before, an interesting situation occurs for the case $q<q'<1$ with $0<a_{q'}<<a_q$, as well as for the 
case $q>q'$ with $a_q<<a_{q'}<0$. For $0\leq x<<x_{crossover}
\equiv [(1-q)a_q]^{\frac{1-q'}{q'-q}}/[(1-q')a_{q'}]^{\frac{1-q}{q'-q}}$ we have 
$y\sim  e_q^{a_qx}$ ($q$-nonextensive statistical mechanics), whereas for 
$x>>x_{crossover}$ we have $y\sim e^{a_{q'}x}_q$ ($q'$-nonextensive statistical mechanics).

Before closing this subsection, let us make a remark concerning the history of physics. It seems quite reasonable to discuss the time dependence of quantities like the sensitivity to initial conditions and the relaxation of physical quantities in terms of a differential equation. But to do so for the energy distribution associated with a stationary state seems, as earlier mentioned, rather amazing. To make such procedure somewhat more acceptable it might be interesting at this point to review a few equations of Planck's October 1900 paper \cite{planck}. He writes the following two equations:
\begin{equation}
\frac{d^2S}{dU^2}=\frac{\alpha}{U(\beta +U)}\ ,
\end{equation}
and
\begin{equation}
\frac{dS}{dU}=\frac{1}{T}\ ,
\end{equation}
where $S$ is the entropy, $U$ the internal energy, $T$ the absolute temperature, 
$\alpha$ and $\beta$ two coefficients to be fixed. The first of these two equations was heuristic. If we replace the second equation into the first one, we obtain
\begin{equation}
\frac{d}{dU}\left(\frac{1}{T}\right)=\frac{\alpha}{U(\beta +U)}
\end{equation}
hence
\begin{equation}
\frac{dU}{d(1/T)}=\frac{\beta}{\alpha}\ U+\frac{1}{\alpha}\ U^2\ .
\end{equation}
It happens that this equation precisely is the $(q',q)=(1,2)$ particular case of our Eq. (77). Planck, in fact, does not write our present Eqs. (82) and (83). After writing our present Eqs. (80) and (81), he jumps (taking into account the density of states) to the writing of the celebrated law for the radiation of the black body. Although not always noticed, the bosonic nature of photons (i.e., the special constraint implied by the symmetrization of the quantum wavefunctions) makes that Planck's law does undergo through a crossover from a power law 
$(q=2)$ at low energies to an exponential law $(q=1)$ at high energies.

\begin{center}
{\bf B. The superstatistics path}
\end{center}

It was noticed in 2000 by Wilk and Wlodarczyk \cite{wilk} that the $q$-exponential distribution can be written as a Laplace transform involving the BG exponential weight. More precisely,
\begin{equation}
e^{-\beta_qE_i}_q=\int^{\infty}_0d\beta \, e^{-\beta E_i}f_q(\beta )\ ,
\end{equation}
where $f_q(\beta )$ is the $\chi^2$ (also called the Gamma) distribution. They further noticed that
\begin{equation}
q=\frac{\langle \beta^2\rangle}{\langle \beta \rangle^2}\ .
\end{equation}
where $\langle \cdots \rangle \equiv \int^{\infty}_0  d\beta \,(\cdots )f_q(\beta )$. This observation was further developed by Beck \cite{beck}. The idea that was emerging was the possible interpretation of nonextensive statistics as a kind of average of BG statistics, where the temperature (or scaled coupling constants) itself would be a stochastic variable. Such mathematical relations were in fact already present in the so called Hilhorst (see \cite{tsallis1994}), 
Prato \cite{prato} and Lenzi \cite{lenzi} formulae, however without any particular physical interpretation.

The next step was recently done by Beck and Cohen when they proposed, through a Laplace transform, their superstatistics 
\cite{beckcohen}. They generalized Eq. (84) into
\begin{equation}
B(E_i)=\int^{\infty}_0d\beta\; e^{-\beta E_i}f(\beta )\ ,
\end{equation}
where $f(\beta)$ is a quite general distribution $(\int^{\infty}_0d\beta f(\beta )=1)$. They consistently generalized Eq. (85) into
\begin{equation}
q_{BC}\equiv \frac{\langle \beta^2\rangle}{\langle \beta \rangle^2}\ ,
\end{equation}
where $\langle \cdots \rangle \equiv \int^{\infty}_0 d\beta (\cdots )f(\beta )$ (the subindex $BC$ stands  for Beck-Cohen and was introduced in \cite{tsallissouza} to avoid con fusion with $q$). Clearly, if $f(\beta )=f_q(\beta )$ we have that $q_{BC}=q$.

Of course, having a weight like that defined in Eq. (86)
is necessary but not sufficient for having a statistical mechanics. We also need an entropy functional, and a consistent manner of writing the constraints such as the energy and similar ones. The optimization of this entropy is expected to be extremal for the superstatistical weight $B(E_i)$. Such an entropy, from now on noted $S_G(\{p_i\})$ ($G$ stands for {\it generalized}), was in fact proposed in \cite{tsallissouza}. It has an univocal relation with 
$f(\beta )$, i.e., one and only one functional $S_G$ corresponds to each admissible 
$f(\beta )$. Of course, $f(\beta )=f_q(\beta )$ implies and is implied by 
$S_G=S_{q}$. This sensibly general entropy $S_G$ is concave by construction. It also happens to be experimentally robust, as proved in \cite{souzatsallis}. Such a convergence of two nontrivial properties (concavity and robustness) is certainly very satisfactory, and constitutes an argument favorable to considering $S_G$ as a physical entropy (i.e., useful for thermal physics), and not just as one more theoretical information measure, among dozens that exist in the literature of cybernetics, control theory, and other applied sciences.

\begin{center}
{\bf C. Connecting the differential-equation and the superstatistics paths}
\end{center}

The solution of Eq. (79) clearly corresponds to a particular $B(E_i)$, from now on noted 
$B_{qq'}(E_i)$, and to its associated $f(\beta )$, from now on noted 
$f_{qq'}(\beta )$. In particular, we have that $B_{q1}(E_i)$ equals the function given in Eq. (78) with $(a_1,a_q,x)\rightarrow  (-\beta_1,-\beta_q,E_i)$. More particularly,
\begin{equation}
B_{21}(E)=\frac{1}{1-\frac{\beta_2}{\beta_1}(1-e^{\beta_1E})}\ .
\end{equation}
If we normalize this function in order to get the probability distribution, take the limit $\beta_2/\beta_1\rightarrow \infty$, and multiply by the d = 3 photonic density of states, we recover, as already discussed, Planck's law.

More details about $B_{qq'}(E_i),\ f_{qq'}(\beta )$, and their application to the recent high-precision experimental (by Bodenschatz et al \cite{bodenschatz}) and computational (by Gotoh et al \cite{gotoh}) results for fully developed turbulence, can be found in \cite{souzatsallis2}. Finally, the logical structure of the successive generalizations of {\it classical} BG statistical mechanics presented in this paper is indicated in Fig. 2. The case of the Planck black-body distribution {\it outside} BG statistical mechanics in the figure deserves a clarification. 
It is well known that, photons being bosons, this distribution naturally belongs to the {\it quantum} version of BG statistical mechanics, more precisely to Bose-Einstein statistics. To be more explicit, {\it classical} BG statistics is contained inside {\it quantum} BG statistics. Bose-Einstein statistics (as well as Fermi-Dirac statistics) is also contained {\it inside} quantum BG statistics though {\it outside} from classical BG statistics. All these distributions (classical BG, Bose-Einstein and Fermi-Dirac)  can obviously be conceived as particular cases of the Beck-Cohen superstatistics since only a Laplace transform is involved (see Eq. (86)). What is special about Planck law is that it can be {\it also} conceived as a simple particular case of the special instance of superstatistics which simultaneously is a solution of Eq. (77).
In other words, the celebrated distribution can also be seen as a crossover between classical BG statistics (i.e., classical $q=1$ statistics) and classical $q=2$ statistics. Summarizing, it is in this {\it specific sense} that this distribution is, as indicated in Fig. 2 , {\it outside} classical BG statistics, and  {\it inside} superstatistics.

\section{CRITIQUES} 

As the history of sciences plethorically shows to us, every possible substantial progress in the foundations of any science is accompanied by controversies. This is a common and convenient mechanism for new ideas to be checked and better understood by the scientific community. There is absolutely no reason to expect that statistical mechanics would be out of it. Quite on the contrary \cite{nicolis}, given the undeniable fact that entropy is one among the most subtle and rich concepts in physics. Indeed, as eloquently commented  by Nicolis and Daems \cite{nicolis}, {\it ``It is the strange privilege of statistical mechanics to stimulate and nourish passionate discussions related to its foundations [...]".} We believe that some space dedicated here to such issues might well be useful at this stage. Therefore, let us address one by one some recent critiques that we are aware of. In fact they have all been replied in specific papers which we indicate here case by case for the interested reader. In what follows, for economy of space, we shall restrict to a brief review of what we understand to be the main focus of each criticism, and what we believe to be the main reason for its rebuttal. \\

\noindent
(a) {\it Vollmayer-Lee and Luijten critique:} \cite{luijten}\\

Vollmayr-Lee and Luijten (VLL) presented in 2001 \cite{luijten} a critique to nonextensive statistical mechanics.  They consider 
a Kac-potential approach of nonintegrable interactions. They consider 
a $d$-dimensional classical fluid with two-body interactions 
exhibiting a hard core as well as an attractive potential 
proportional to $r^{-\alpha}$ with $0 \le \alpha/d < 1$ (logarithmic 
dependence for $\alpha/d = 1$; VLL use the notation $\tau \equiv \alpha$). In their approach, they 
also include a Kac-like long-distance cutoff $R$ such that no 
interactions exist for $r>R$, and then discuss the $R \rightarrow 
\infty$ limit. They show that the exact solution within 
Boltzmann-Gibbs statistical mechanics is possible and that -- no 
surprise (see VLL Ref. [12] and references therein) -- it exhibits a 
mean field criticality. Moreover, the authors argue that very similar 
considerations hold for lattice gases, $O(n)$ and Potts models. 

VLL state {\it ``Our findings imply that, contrary to 
some claims, Boltzmann-Gibbs statistics is sufficient for a standard 
description of this class of nonintegrable interactions."}, and also 
that {\it ``we show that nonintegrable interactions do not require 
the application of generalized $q$-statistics."}. 
In our opinion, these statements may misguide the reader. The critique was rebutted in \cite{commentluijten}, whose main points are summarized here. Indeed, the 
VLL discussion, along traditional lines, of their 
specific Kac-like model only exhibits that Boltzmann-Gibbs 
statistical mechanics is -- as more than one century of brilliant 
successes guarantees! -- {\it necessary} for calculating, {\it without doing time averages}, a variety of {\it thermal 
equilibrium} properties; by 
no means it proves that it is {\it sufficient}, as we shall soon 
clarify. Neither it proves that wider approaches such as, for 
instance, nonextensive statistical mechanics (VLL Refs. [6,31] and present \cite{tsallis,tsallis2,tsallis3}), or any other similar formalism that might emerge, are not required 
or convenient. The crucial point concerns {\it time}, a word that 
nowhere appears in the VLL paper. The key role of $t$ has been emphasized in several 
occasions, for instance in Fig. 4 of VLL Ref. [31]. For integrable or short-range interactions (i.e., for $\alpha/d > 1$), we expect that 
the $t \rightarrow \infty$ and $N \rightarrow \infty$ limits are 
commutable  in what concerns the equilibrium distribution $p(E)$, $E$ 
being the total energy level associated with the macroscopic system. 
More precisely, we expect naturally that (excepting for the density of states)
\begin{eqnarray}
p(E) &\equiv& \lim_{t \rightarrow \infty} \lim_{N \rightarrow \infty} 
p(E;N;t) 
= \lim_{N \rightarrow \infty} \lim_{t \rightarrow \infty} p(E;N;t)  
\nonumber \\
&\propto& \exp[-E/kT]  \;\;\;(\tau/d >1)
\end{eqnarray} 
if the system is in thermal equilibrium with a thermostat at temperature 
$T$. In contrast, the system is expected to behave in a more complex 
manner for nonintegrable (or long-range) interactions, i.e., for 
$0 \le \alpha/d \le 1$. In this case, no generic reason seem to exist 
for the $t \rightarrow \infty$ and $N \rightarrow \infty$ limits to 
be commutable, and consistently we expect not necessarily equal 
results. The simplest of these results (which is in fact the one to 
be associated with the VLL paper, although therein these two relevant 
limits and their ordering are not mentioned) is, as we shall soon 
further comment,
\begin{equation}
\lim_{N \rightarrow \infty} \lim_{t \rightarrow \infty}    p(E;N;t) 
\propto \exp[-(E/{\tilde N})/(kT/{\tilde N})] \;.
\end{equation} 
${\tilde N} \equiv 
[N^{1-\alpha/d}-\alpha/d]/[1-\alpha/d]$ has been introduced in order to stress the facts that 
{\it generically} 

(i) $E$ is {\it not} extensive, i.e., is {\it not} proportional to 
$N$ but is instead $E \propto N{\tilde N}$  [more precisely, $E$ is 
extensive if $\alpha/d >1$ (see \cite{fisher} and VLL Refs. [4,5]), and 
it is nonextensive if $0 \le \alpha/d \le 1$];
and 

(ii) $T$ needs, in such calculation, to be rescaled (a feature which is frequently absorbed 
in the literature by artificially size-rescaling the coupling 
constants of the Hamiltonian), in order to guarantee nontrivial {\it 
finite} equations of states. Of course, for $\alpha = 0$, we have 
${\tilde N}=N$, which recovers the traditional Mean Field scaling. 

But, {\it depending on the initial conditions}, which determine the 
time evolution of the system if it is assumed isolated, quite {\it 
different} results can be obtained for the ordering $\lim_{t 
\rightarrow \infty} \lim_{N \rightarrow \infty}    p(E;N;t)$. This 
fact has been profusely detected and stressed in the related 
literature (see, for instance, VLL Ref. [31], present Refs. 
\cite{antoniruffo,vitoandreaconstantino,aging,cabral,spinglass,posch,antonitorcini,ruffoetal} 
and references therein). Unfortunately, this important fact has been missed in the VLL critique. Such metastable states can {\it by no means} be described within BG statistical mechanics. Even more, as shown in Section IV, there is nowadays increasing evidence that they might be intimately related to nonextensive statistical mechanics. In any case, it is plain that, for such long-range Hamiltonians, BG statistics is necessary but not sufficient, in contrast with the VLL statements. \\

\noindent
(b) {\it Nauenberg critique:} \cite{nauenberg}\\

Some line of critique concerns whether the zeroth principle of thermodynamics and thermometry are consistent with nonextensive statistical mechanics. Such questioning is by no means new: a couple of dozens of papers are available in the literature which address this important point.  It has been recently raised once again, this time by Nauenberg \cite{nauenberg}. He concludes, among many other critiques, that it is not possible to have thermalization between systems with different values of $q$. It appears to be exactly the opposite which is {\it factually} shown in \cite{tsallisreply}. His critique is rebutted in \cite{tsallisreply}.  One of the crucial points that is unfortunately missed in \cite{nauenberg}, concerns discussion of ``weak coupling" in Hamiltonian systems.  Indeed, if we call $c$ the coupling constant associated with long range interactions (i.e., $0 \le \alpha/d \le1$), we have that $ \lim_{N \to \infty} \lim_{c \to 0} c \tilde N =0$, whereas $ \lim_{c \to 0}\lim_{N \to \infty}  c \tilde N$ diverges. No such anomaly exists for short-range interactions (i.e., $\alpha/d>1$). Indeed, in this simpler case, we have that $ \lim_{N \to \infty} \lim_{c \to 0} c \tilde N = \lim_{c \to 0}\lim_{N \to \infty}  c \tilde N=0$. The nonuniform convergence that, for long-range interactions, exists at this level  possibly is related to the concomitant nonuniform convergence associated with the $t \to \infty$ and $N \to \infty$ limits discussed previously in this paper. All these subtleties are not mentioned in \cite{nauenberg}. \\

\noindent
(c) {\it Luzzi, Vasconcellos and Ramos critique:} \cite{luzzi}\\

Another line of critique concerns the ``physicality" of $S_q$ (see \cite{luzzi}). Or whether it could exist a ``physical" entropy different from $S_{BG}$. Since such issues appear to be of a rather discursive/philosophical nature, we prefer to put these critiques on slightly different, more objective, grounds. We prefer to ask, for instance, (i) whether $S_q$ is useful in theoretical physics in a sense similar to that in which $S_{BG}$ undoubtedly is useful; (ii) whether $q$ necessarily is a fitting parameter, or whether it can be determined a priori, as it should if we wish the present theory to be a complete one; (iii) whether there is no other way of addressing the thermal physics of the anomalous systems addressed here, very specifically whether one could not do so by just using $S_{BG}$; (iv) whether $S_q$ is special in some physical sense, or whether it is to be put on the same grounds as the twenty or  thirty entropic functionals popular in cybernetics, control theory, and information theory generally speaking. Such questions have received answers in \cite{abepreprint,tsalliscordoba,creta,science1,science2,science3} and elsewhere.  (i) The usefulness of this theory seems to be answered by the large amount of applications it has already received, and by the ubiquity of the $q$-exponential form in nature. (ii) The a priori calculation of $q$ from microscopic dynamics has been specifically illustrated in Section III. (iii) The optimization of $S_q$, as well as of almost any other entropic form, with a few constraints has been shown in \cite{tsalliscordoba}
to be equivalent to the optimization of $S_{BG}$ with an infinite number of appropriately chosen constraints. Therefore we could restrain to the use of $S_{BG}$ if we absolutely wanted that, similarly to the fact that, instead of using the extremely convenient Keplerian ellipse for the planetary orbits, we could equivalently use an infinite number of Ptolemaic epicycles. It is however appreciably much simpler to represent a complex structure of constraints into the single index $q \ne 1$ (in analogy with the fact that the ellipticity of a Keplerian orbit can be simply specified by a single parameter, namely the eccentricity of the ellipse).  (iv) The entropy $S_q$ shares with $S_{BG}$  an impressive set of important properties, which namely includes concavity, stability and finiteness of the entropy production per unit time, $\forall q>0$. The difficulty of simultaneously satisfying all these three properties can be measured by the fact that the Renyi entropy (usefully used in the geometric characterization of multifractals) satisfies none of them for all $q>0$. Such features point $S_q$ as being very special for thermostatistical purposes.  \\

\noindent
(d) {\it Zanette and Montemurro critique I:} \cite{zanettecritique1}\\

In a recent paper \cite{zanettecritique1}, Zanette and Montemurro re-analyze the molecular dynamics approach and results presented in \cite{vitoandreaconstantino} for the infinitely-long-range interacting planar rotators already discussed here. They especially focus on the time dependence of the temperature $T(t)$ defined as the mean kinetic energy per particle. For total energy slightly below the second-order critical point and a non-zero-measure class of initial conditions, a long-standing nonequilibrium state emerges before the system achieves the terminal BG thermal equilibrium. When $T(t)$ is plotted, as done by virtually all authors, by using a linear scale for $T$ and a logarithmic scale for $t$, an inflection point exists. If we call $t_{crossover}$ the value of $t$ at which the inflection point is located, it has been repeatedly verified numerically by various authors, including Zanette and Montemurro \cite{zanettecritique1}, that $\lim_{N \to\infty}t_{crossover}(N)$ diverges. Therefore, if the system is very large (in the limit $N\to\infty$, mathematically speaking) it remains virtually for ever in the anomalous state, currently called by many authors, {\it quasi-stationary state} or {\it metastable state}. Zanette and Montemurro point out (correctly) that, if a linear scale is used for $t$, the inflection point disappears. From this, these authors conclude that this well known metastable state is but a kind of mathematical artifact, and no physically relevant quasi-stationarity exists. It is like if the high-to-low energies crossing occurring, at a given temperature, in Fermi-Dirac statistics had no physical meaning! Indeed, if instead of using the linear scale for the energies we were to use a faster scale (e.g., an exponential scale), the well known inflection point will surely disappear. Nevertheless, there is no point to conclude from this that the textbook crossing in Fermi-Dirac statistics is but a mathematical artifact. In fact, {\it any} inflection point on {\it any} curve will disappear by sufficiently ``accelerating" the abscissa. Coming back to the system of rotators, what indeed appears to happen is that, for increasingly large $N$, $T(t)$ remains constant, and {\it different from the BG value}, within an error bar which appears to {\it vanish} in the $N \to\infty$ limit. This effect appears in an even more pronounced way because of a slight minimum that $T(t)$ presents just before going up to the BG value. This intriguing minimum had already been observed in \cite{vitoandreaconstantino} and has been detected with higher precision in  \cite{zanettecritique1}. A partial rebuttal of this critique is presented in \cite{spinglass}.\\

\noindent
(e) {\it Zanette and Montemurro critique II:} \cite{zanettecritique2}\\

Soon after the previous one, Zanette and Montemurro advanced a second critique \cite{zanettecritique2} objecting the validity of nonextensive statistical mechanics for thermodynamical systems. This line of critique concerns the ubiquity of the $q$-exponential form as a stable law in nature. The argument essentially goes that only Gaussians and L\'evy distributions would be admissible, because of the respective central limit theorems. Such question has been addressed long ago in \cite{bologna} (ignored in \cite{zanettecritique2}), and once again more recently in \cite{economics}, as a rebuttal to \cite{zanettecritique2}. The answer basically reminds that the time stability of the distribution only has a mathematically well defined meaning if we {\it also} specify the time evolution or time composition law. The just mentioned central limit theorems {\it only} apply for convolution time evolution, certainly not for nonmarkovian evolutions as those illustrated in Section IV.  \\

\noindent
(f) {\it Zanette and Montemurro critique III:} \cite{zanettecritique3}\\

Soon after the second critique, Zanette and Montemurro advanced a third one \cite{zanettecritique3}. This time the objection addresses the rest of the systems, i.e., the non thermodynamical ones. 
It is argued by these authors that non thermodynamical applications of nonextensive statistics are ill-defined, essentially because of the fact that {\it any} probability distribution can be obtained from the nonextensive entropy $S_q$ by conveniently adjusting the constraint used in the optimization. We argue here that, since it is well known to be so for  {\it any} entropic form and, in particular, for the Boltzmann-Gibbs entropy $S_{BG}$, the critique brings absolutely no novelty to the area. In other words, it has {\it nothing} special to do with the nonextensive entropy $S_q$. In defense of the usual simple constraints, typically averages of the random variable $x_i$ or of $x_i^2$ (where $x_i$ is to be identified according to the nature of the system), we argue, {\it and this for all entropic forms}, that they can hardly be considered as arbitrary, as  Zanette and Montemurro seem to consider. Indeed, once the natural variables of the system have been identified, the variable itself and its square obviously are the most basic quantities to be constrained. Such constraints are used in hundreds (perhaps thousands) of useful applications outside (and also inside) thermodynamical systems, along the information theory lines of Jaynes and Shannon, and more recently of A. Plastino and others. And this is so for $S_{BG}$, $S_q$, and any other entropic form. If, however, other quantities are constrained (e.g., an average of $x^\sigma$ or of $|x|^\sigma$) for specific applications, it is clear that, at the present state-of-the-art of information theory, {\it and for all entropic forms},  this must be discussed case by case. The full rebuttal of this critique can be found in \cite{replyzanettecritique3}. \\

\noindent
(g) {\it About ordinary differential equations:}\\
 
The remarks in Sections I.A and II.A related to ordinary differential equations might surprise some readers, hence deserve some clarification. Indeed, in virtually all the textbooks of statistical mechanics, functions such as the energy distribution at thermal equilibrium are discussed {\it using a  variational principle}, namely referring to the entropy functional, and  {\it not using ordinary differential equations} and their solutions. In our opinion, it is so {\it not} because of some basic (and unknown) principle of exclusivity, but rather because the first-principle dynamical origin of the BG factor still remains, mathematically speaking, at the status of a {\it dogma} \cite{takens}. Indeed, to the best of our knowledge, no theorem yet exists which establishes the necessary and sufficient conditions for being valid the use of the celebrated BG factor. Nevertheless, one must not forget that it was precisely through a differential equation that Planck heuristically found, as described in his famous October 1900 paper \cite{planck}, the black-body radiation law. It was only in his equally famous December 1900 paper that he made the junction with the Boltzmann factor by assuming the --- at the time, totally bizarre --- hypothesis of discretized energies.

A further point which deserves clarification is {\it why} have we {\it also} interpreted the linear ordinary differential equation in Section I.A as providing the typical time evolution of both the sensitivity to the initial conditions and the relaxation of relevant quantities. Although the bridging was initiated by Krylov \cite{krylov}, the situation still is far from  completely clear on mathematical grounds. However, intuitively speaking, it seems quite natural to think that the sensitivity to the initial conditions is precisely what makes the system to relax to equilibrium, and therefore opens the door for the BG factor to be valid. In any case, although some of the statements in Section I.A are (yet) not proved, this by no means implies that they are generically false. Furthermore, they provide what we believe to be a powerful metaphor for generalizing the whole scheme into the nonlinear ordinary differential equations discussed in Section II.A. Interestingly enough, the $q$-exponential functions thus obtained have indeed proved to be the correct answers for a sensible variety of specific situations reviewed in the rest of the present paper, and this for {\it all three} interpretations as energy distribution for the stationary state, time evolution of the sensitivity to the initial conditions, and time evolution of basic relaxation functions. \\    

\section{FINAL COMMENTS}

We all know that the concepts of energy and entropy are cornerstones of contemporary physics. They are both at the heart of thermodynamics, a set of connections that regulate the laws of the macroscopic world. Statistical mechanics bridges microscopic dynamics (Newtonian, quantum, relativistic, or any other) with the macroscopic behaviors we observe in nature. Boltzmann-Gibbs statistical mechanics constitutes an impressive illustration of how the logics of these micro-macro connections is structured. This theory is based on the BG expression $S_{BG}$ for the entropy. Following along the lines of Einstein 1910 and of many others, it is our belief that the specific mathematical form that    $S_{BG}$ has must ultimately descend from microscopic dynamics. More precisely, this connection is known to rely on quick mixing in phase space, ergodicity, and related nonlinear dynamical concepts. This is the statistical mechanics of short range interactions, short-range microscopic memory, ultimately leading to a simple, uniform  occupancy of the allowed phase space (the ``equal probability" hypothesis for the microcanonical ensemble). The question arises naturally: what happens when the interactions are long-ranged, when the microscopic memory is heavily nonmarkovian, when the geometry of occupation of phase space is complex, (multi)fractal for instance, or some other hierarchical structure? Can we still devise theoretical techniques similar to those of BG statistical mechanics, which would once again bridge with the laws of the macroscopic world? We believe that we can. A central point therefore is: what mathematical expression should or could we use for the entropy? How can we adequately generalize $S_{BG}$?     

The answer to such question obviously is far from trivial. Indeed, even in the framework of BG statistical mechanics, we do not know how to rigorously make the expression of $S_{BG}$ descend from microscopic dynamics, plus possibly some other generic logical requests. In some sense, the best we know nowadays about such fundamental question lies at the level of the necessary and sufficient conditions proposed by Shannon, by Khinchin, and similar constructs. {\it But} they do not start from dynamics.

Since the direct connection between normal (ergodic) microscopic dynamics and $S_{BG}$ is yet not totally clear, there is no surprise that the same happens for anomalous microscopic dynamics. This is the fundamental reason for which we have proceeded to generalize $S_{BG}$ into $S_q$ through a metaphoric path, as illustrated in Section II. The $q$-generalizations of the Shannon and the Khinchin theorems have already been established by Santos and by Abe respectively. A variety of arguments qualify  the conditions under which $S_q$ is unique (see \cite{abepreprint} and references therein). Still, how to make $S_q$ (or even $S_{BG}$, as we said before) descend from microscopic dynamics remains an open question. Consistently remains an open question what exactly have in common all the physical cases indicated in Section III, and what may exactly be the geometrical structure which reflects the dynamical occupancy of the phase space in such cases. We have conjectured that it may well be similar to a scale-free geometry like that of the Albert-Barabasi model, but this remains to be proved. 

Another intriguing question is the connection with aging, a property that long-range-interacting Hamiltonians share with many glasses, spin-glasses, and other metastable systems intensively studied nowadays. Such long-range Hamiltonians satisfy, even in the longstanding aging phase (whose duration diverges with $N$), a zeroth principle of thermodynamics, which concretely opens the path to thermodynamics. Furthermore, a sensible amount of connections with nonextensive statistical mechanics are already available in the literature. However, the exact dependence of $q$ on $(\alpha,d)$, possibly on $\alpha/d$, is still to be unambiguously established. Many researchers around the world are presently working on that fascinating problem.    

Finally, we should emphasize that everything that we know today neatly points towards the scenario that nonextensive statistical mechanics satisfies the $0$-th, first, second and third principle of thermodynamics (see, for instance, \cite{aberajagopalsecond}). This is deeply interesting, since that implies that the basic laws of thermodynamics are stronger than the role reserved for them within Boltzmann-Gibbs statistical mechanics.     
Further analysis of the foundations and thermodynamical connections of statistical mechanics in general, and of nonextensive statistical mechanics in particular, are certainly fascinating and very welcome.   
   
\section{ACKNOWLEDGMENTS}

I have in more than one occasion referred in this paper to 1900 Planck's contributions. This was in fact induced by delightful conversations I had with L. Tisza in 1995 (in H.E. Stanley's office at the Boston University), and with L. Galgani in 2001 (during a Les Houches Winter School organized by T. Dauxois, S. Rufo, E. Arimondo and M. Wilkens). I warmly acknowledge here this double privilege.

I have also benefited from many invaluable discussions with M. Gell-Mann, as well as from many and interesting remarks from S. Abe, G.F.J. Ananos, C. Anteneodo, F. Baldovin,  M. Baranger, E.P. Borges, E. Brigatti, M.O. Caceres, E.G.D. Cohen, E.M.F. Curado, L.G. Moyano, F.D. Nobre, A. Plastino, A.R. Plastino, A.K. Rajagopal, A. Rapisarda and A.M.C. Souza. Moreover, I am grateful to three anonymous Referees, whose helpful comments have resulted in the present, hopefully improved, version of the original manuscript. 

Finally, without the intense and generous interest of H.L. Swinney and A.R. Bishop, and others, in organizing and supporting the Los Alamos National Laboratory {\it International Workshop on Anomalous Distributions, Nonlinear Dynamics and Nonextensivity}, the present volume would have not existed.

\begin{figure}
\begin{center}
\includegraphics[width=9.0cm,angle=0]{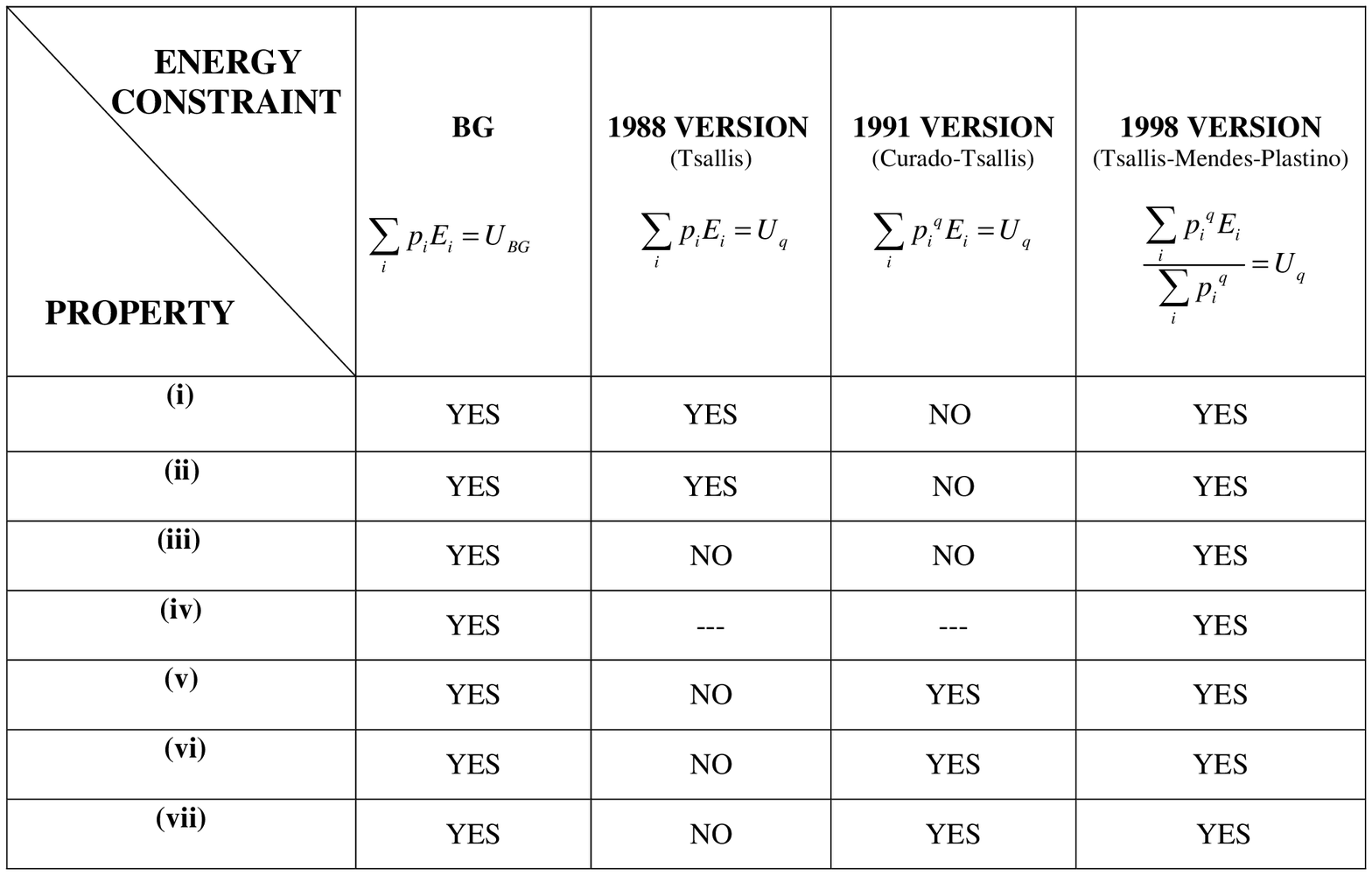}
\end{center}
\caption{\small Successive versions for the canonical-ensemble energy-constraint within nonextensive statistical mechanics regarding properties (i-vii) indicated in Section III assuming fixed the Lagrange parameter $\beta$. In the 1988 paper [12]  two possible constraints for the energy were indicated , namely $\sum_i p_i E_i = constant$ and $\sum_i p_i^q E_i = constant$, but only the former was developed therein. In the 1991 paper [13], the second form for the constraint was developed, which enabled the connection with thermodynamics. In the 1998 paper [14], the constraint using escort distributions was adopted and developed, i.e., $\sum_i p_i^q E_i/ \sum_i p_i^q = constant$. This last form provides for the points (i-vii) the {\it same} answers as within BG statistical mechanics, and is presently believed to be its correct $q$-generalization.} 
\end{figure}

\begin{figure}
\begin{center}
\includegraphics[width=7.5cm,angle=0]{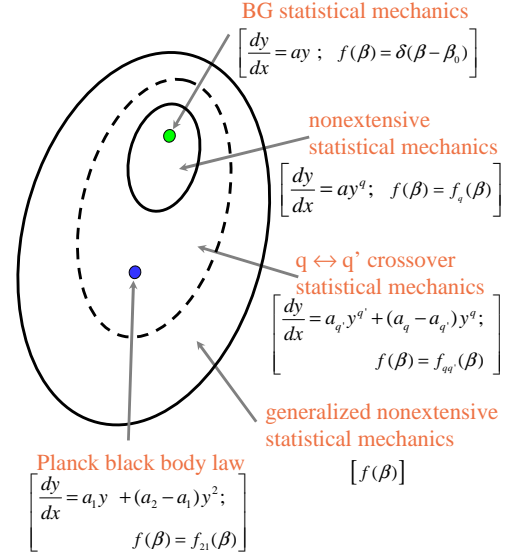}
\end{center}
\caption{\small Logical structure of the successive generalizations 
of {\it classical} BG statistical
mechanics presented in this paper, with 
indication of their  paradigmatic differential equations, and their $f(\beta )$ superstatistical distributions; 
$f_1(\beta )=f_{BG}(\beta )\ ;  f_q(\beta )=\chi^2$ -distribution. The corresponding stationary state probability distributions and entropies are given by Eqs. (9) and (10), for BG statistical mechanics, by Eqs. (30) and (34), for nonextensive statistical mechanics, and by Eq. (86) and $S_G$ given in [68,69],  for generalized nonextensive statistical mechanics. The $q \leftrightarrow q^\prime$ crossover statistical mechanics includes as a particular case the $q \leftrightarrow 1$ crossover one, which corresponds to Eq. (88). If we take into account in this equation the normalization factor, the $\beta_2/\beta_1 \to \infty$ limit, and the photonic density of states, we recover Planck's law for the black-body radiation. See the text in Section V.C for clarification of this case.} 
\end{figure}

\end{multicols}

\end{document}